\documentclass[a4paper]{IEEEtran}

\usepackage[percent]{overpic}
\usepackage{array}
\usepackage{epstopdf}
\usepackage{fancyhdr}

\usepackage{lmodern}
\usepackage{textcomp}
\usepackage{algorithmic}
\usepackage{bigstrut}
\usepackage{rotating}
\usepackage{tabularx}
\usepackage{pifont}
\usepackage{multirow}

\usepackage{graphics,graphicx,amsopn,amsmath,amssymb,amsfonts}
\usepackage{cite}
\usepackage{citesort}

\usepackage{colortbl}

\definecolor{greycolor1}{cmyk}{0,0,0,1}

\newcommand{\cmark}{\ding{51}}%
\newcommand{\xmark}{\ding{55}}%

\newcolumntype{L}[1]{>{\raggedright\let\newline\\\arraybackslash\hspace{0pt}}m{#1}}
\newcolumntype{C}[1]{>{\centering\let\newline\\\arraybackslash\hspace{0pt}}m{#1}}
\newcolumntype{R}[1]{>{\raggedleft\let\newline\\\arraybackslash\hspace{0pt}}m{#1}}

\def\BibTeX{{\rm B\kern-.05em{\sc i\kern-.025em b}\kern-.08em
    T\kern-.1667em\lower.7ex\hbox{E}\kern-.125emX}}

\begin{document}

\title{\Large\bf The Potential Short- and Long-Term Disruptions and Transformative Impacts of 5G and Beyond Wireless Networks: Lessons Learnt from the Development of a 5G Testbed Environment}

\author{
Mohmammad N. Patwary, Syed Junaid Nawaz, Md. Abdur Rahman, Shree Krishna Sharma, \\
Md Mamunur Rashid, Stuart J. Barnes
\thanks{M. N. Patwary is with the School of Computing and Digital Technology, Birmingham City University, Birmingham, UK. (e-mail: {\it mohammad.patwary@bcu.ac.uk})}
\thanks{S. J. Nawaz is with the Department of Electrical and Computer Engineering, COMSATS University Islamabad (CUI), Islamabad 45550, Pakistan. (e-mail: {\it junaidnawaz@ieee.org})}
\thanks{ M. A. Rahman is with the Department of Cyber Security and Forensic Computing, The University of Prince Mugrin, KSA. (e-mail: {\it m.arahman@upm.edu.sa})}
\thanks{S. K. Sharma is with the SnT - securityandtrust.lu, University of Luxembourg, Kirchberg, Luxembourg 1855, Luxembourg. (e-mail: {\it shree.sharma@uni.lu})}
\thanks{M. M. Rashid and S. J. Barnes are with the Consumer and Organisational Data Analytics (CODA) Research Centre, King's College London, UK. (e-mail: {\it mamun.rashid@kcl.ac.uk} and {\it stuart.barnes@kcl.ac.uk} )}
\thanks{Corresponding author: Syed Junaid Nawaz (e-mail: {\it junaidnawaz@ieee.org})}
\thanks{\copyright 2020 IEEE. Personal use of this material is permitted. Permission from IEEE must be obtained for all other uses, in any current or future media, including reprinting/republishing this material for advertising or promotional purposes, creating new collective works, for resale or redistribution to servers or lists, or reuse of any copyrighted component of this work in other works}
}

\markboth{DOI: 10.1109/ACCESS.2020.2964673}{}
\maketitle

\begin{abstract}
The capacity and coverage requirements for 5G and beyond wireless connectivity will be significantly different from the predecessor networks. To meet these requirements, the anticipated deployment cost in the UK is predicted to be in between \textsterling30bn- \textsterling50bn, whereas the current annual capital expenditure (CapEX) of the mobile network operators (MNOs) is \textsterling2.5bn. This prospect has vastly impacted and has become one of the major delaying factors for building the 5G physical infrastructure, whereas other areas of 5G developments are progressing at their speed. Due to the expensive and complicated nature of the physical network infrastructure and spectrum, the second-tier operators, widely known as mobile virtual network operators (MVNO), are entirely dependent on the MNOs. In this paper, an extensive study is conducted to explore the possibilities of reducing the 5G deployment cost and developing business models. This study suggests that the use of existing public infrastructure (e.g., streetlights, telephone poles, etc.) has a great potential to contribute to a reduction of about 40\% to 60\% in the anticipated cost. This paper also reviews the recent Ofcom initiatives to release location-based licenses of the 5G-compatible radio spectrum at a nominal cost. Our study suggests that simplification of infrastructure and spectrum will encourage the exponential growth of scenario-specific cellular networks and will potentially disrupt the current business models of telecommunication business stakeholders -- specifically MNOs and TowerCos. These scenario-specific networks are expected to be: a) private networks, b) community networks, and c) micro-operators. Furthermore, due to the feasibility of dense device connectivity with 5G, the resolution of traditional and non-traditional data availability will increase significantly. This will encourage extensive data harvesting as a business opportunity and function within small and medium-sized enterprises (SMEs) as well as within large social networks. Consequently, the rise of new infrastructures and spectrum stakeholders is anticipated. This will fuel the development of a 5G data exchange ecosystem where data transactions are deemed to be high-value business commodities. The privacy and security of such data, as well as definitions of the associated revenue models and ownership, are challenging areas -- and these have yet to emerge and mature fully. In this direction, this paper proposes the development of a unified data hub with layered structured privacy and security along with blockchain and encrypted off-chain based ownership/royalty tracking. Also, a data economy-oriented business model is proposed. The study found that with the potential commodification of data and data transactions along with the low-cost physical infrastructure and spectrum, the 5G network will introduce significant disruption in the Telco business ecosystem.
\end{abstract}
\smallskip
\textbf{keywords:} 5G, 5G Deployment, 5G Testbed, B5G, Infrastructure Sharing, Security, Teleco Business

\section{Introduction}

The rollout of 5\textsuperscript{th} generation (5G) of communication networks has commenced with Release-15 of 3\textsuperscript{rd} Generation Partnership Project (3GPP) \cite{3GPPrel15}. This release of 5G new radio (5G NR) has extended the provisions for both standalone and non-standalone operations. Based on the ongoing field trials and non-commercial (test) deployment based investigations, the standardization of 5G is expected to mature with Release-16 of 3GPP by the year 2020.
The 5G networks are expected to bring a transformative impact in the role that mobile communication technologies play in the society \cite{obiodu20175g}. The 5G has taken a huge leap forward in the offered services, which are introduced through the advent of various new innovative technologies.
The notable target 5G services can be named as enhanced mobile broadband (eMBB), ultra-reliable low latency communications (URLLCs),  massive machine-type communications (mMTCs), and Tactile Internet (TI) \cite{7414384,shafi20175g,Simsek2016tactile}. To solely benefit from the services offered by 5G technologies, the substantiation of 5G affordability and business case is a vital necessity.

\subsection{Motivation for 5G Business Case}

Despite the revolutionary technologies and innovative services being offered by 5G networks, the difficulties associated with their deployment along with other technological shortcomings have already started to appear in the literature, see e.g.,  \cite{schneir2019business,wisely2018capacity,8412482,chianiopen}. The registered critical limitations and challenges for 5G networks can be summarized as: i) the enormous deployment cost, ii) the explosion of connected devices caused by the advent of mMTCs may very rapidly lead towards reaching the network capacity limit, iii) the rate and volume of the data generated in hyper massively connected 5G networks may need new data analytic innovations, and iv) the privacy and security provisions in massively connected networks -- to name a few.
The telecommunication engineers, industries, and researchers from around the globe have also already initiated the speculative propositions for network requirements and candidate technologies for beyond 5G (B5G) networks, see e.g.,  \cite{QML_6G_Junaid,Saad_6G_2019,Tariq6G_2019,8760275,8782879}.

With a drastic increase in mobile internet users, the subscribers are likely to reach 5.0bn over the next 15 years \cite{Report_MobileEconomy2019}. Moreover, the sole mobile subscribers are expected to rise to 5.8bn between the years 2018 to 2025. A contribution of \$ 2.2 trillion from 5G technologies to the global economy in the next 15 years is projected \cite{Report_MobileEconomy2019}. To facilitate this generation shift, a capital expenditure (CapEx) of \$ 480bn is anticipated from the MNOs between the years 2018 and 2020. Moreover, with most of the 5G services happening after 2020, the CapEx will significantly exceed the CapEx anticipated by 2020.
Admittedly, a huge investment for the network infrastructure deployment is a prerequisite to fully reap the benefits offered by the 5G networks. This necessitates the development of a comprehensive business model for 5G rollout to convince the investors, mobile operators, and other stake-holders to invest the requisite revenue. To this end, very recently, only a few articles discussing the 5G rollout cost, business cases, and other associated implications have appeared in the literature. Nevertheless, there is a strong need to thoroughly research the 5G rollout implications and to develop the sustainable business models addressing the complete scope of all 5G services (e.g., eMBB, mMTC, TI, etc) and all network scenarios (e.g., urban, rural, etc).

\subsection{Potential Solutions for 5G Affordability}

The affordability of delivering the eMBB services offered by 5G is regarded as a vital issue by International Telecommunication Union (ITU) broadband commission \cite{8300495}. The potential domains that can be explored to meet the challenge of making the ultra-fast mobile broadband as affordable to further enable the smooth provision of different societal and other interesting services can be identified as: i) infrastructure sharing, ii) Neutral hosting, iii) unlicensed spectrum utilization, iv) location-based spectrum licensing, v) energy-efficient networking, vi) wireless backhauling, vii) infrastructure cost reduction through softwarization and virtualization of the network functions, and viii) data sharing.
Furthermore, in \cite{8516977}, a comparison of different potential technologies for the provision of network backhaul has been conducted. The fixed wireless backhaul is suggested as the most cost-effective solution for high cost fibers, while the utilization of low cost fibers has been suggested to result in direct-fiber technology as the most cost-effective solution. In \cite{oughton2019assessing}, the 5G infrastructure strategies for capacity, coverage, and cost of 5G eMBB have been discussed for the Netherlands. The analysis for both supply-driven and demand-driven investment has been conducted. Also, the potential for traffic capacity enhancement in the existing Dutch macrocell network with the integration of the new 5G spectrum (only) has been studied. It has been determined that the average per-user capacity enhancement of 40\%, as compared to the existing 4\textsuperscript{th} generation (4G) networks, can be achieved by solely integrating the new 5G spectrum (i.e., without deploying the small-cells). The further improvement required beyond this determined threshold will necessitate the densification of the network with small-cells.
The development of a framework to study the affordability of all 5G services, beyond eMBB, is a vital need at the current context.
Moreover, the establishment of worldwide prospective on the affordability of 5G rollout is another important open challenge.

\subsection{5G Rollout in the United Kingdom}
In this section, we discuss about various aspects of 5G rollout in the United Kingdom (UK) including the recent developments. The ambitions of the UK  government to be a global leader in 5G communication network technology needs the resolution of barriers and challenges in the commercial deployment of 5G. In this direction,  the article \cite{jones2019commentary} provided a review of critical problems related to the market factors for 5G rollout in the UK.
In \cite{yaghoubi2019techno}, a framework for the techno-economic market analysis of network backhaul has been proposed. The proposed framework can be used to study the total cost of ownership (TCO) of a network backhaul and business feasibility of the 5G network deployment aspects. The module takes the consideration of both the network CapEx and OpEx. A case study has also been conducted to demonstrate the usability of the proposed framework. A thorough analysis of techno-economic aspects of deploying eMBB in a typical dense urban area, realized by a 1km$^2$ grid of central London in the UK, has been conducted in \cite{wisely2018capacity}. Various aspects such as CapEx/OpEx, maximum user rate, and capacity have been studied for macro, micro, and hot-spot cellular network settings at 700MHz, 3.5GHz, and 24--27.5GHz operating frequencies, respectively. The headline rate of 64 to 100 Mbps, across all the coverage area, is expected to be achieved through several different technology prospects. However, the use of mmWave and 802.11ac are advised as necessary for achieving capacity in the orders of 100Gbps/km$^2$ for outdoor and indoor settings, respectively. It has been speculated that a 100-fold increased capacity and 100Mbps headline rate everywhere may be attained with an escalation of 4 to 5 times in the deployment cost as compared to that of 4G Long Term Evolution (LTE) networks.

Furthermore, in \cite{schneir2019business}, a business model to study the cost and revenue flow of 5G has been proposed by conducting the case studies for three different boroughs of central London, UK. The eMBB services of 5G are considered important for the business case between the years 2020 to 2030. Some business risks that may emerge in the later years are also indicated. The network share has been highlighted as a significantly helpful aspect in improving the business case. Moreover, it is encouraged to conduct further research for different regions to obtain a nationwide understanding of the business case.  In \cite{oughton2018cost}, the 5G rollout implications in UK have been discussed. The history of implications faced in the 4G rollout has been extrapolated to forecast the characteristics between the years 2020 and 2030. It has been concluded that the 5G eMBB may reach out to 90\% of Britain's population by 2027.
The challenges associated with capital intensity fluctuations may affect the pace of 5G rollout to the rural areas. Some infrastructure sharing suggestions for deploying small-cells can be considered to reduce the 5G deployment cost.  The Ofcom has recently initiated the location-based licensing of 5G compatible radio spectrum \cite{ofcom_licence}, which will significantly help in making 5G affordable in the UK.  Moreover,  authors in \cite{oughton2018cost} thoroughly discussed the policy matters and potential directions to drive the 5G rollout. Furthermore, the need for incorporating more spectrum for serving ultra-high-speed broadband to rural area users has also been suggested. In summary, the development of a comprehensive understanding of the disruptive impacts of 5G and beyond wireless networks in the UK is of vital importance in promoting a sustainable and ambitious digital economy in the long-term.

\subsection{Contributions and Organization}

The main contribution of this paper is the investigation of potential long-  and  short-term transformative and disruptive impacts of 5G rollout. Potential solutions for reducing the 5G deployment cost and developing long-term sustainable business model for 5G through network infrastructure sharing, public infrastructure sharing, radio spectrum sharing, and data sharing are proposed. A typical UK city is taken for conducting the case study in this paper. Different to the approach of extrapolating the historic prospective adopted in \cite{oughton2018cost}, our work intends to provide the prospective learnt from a 5G testbed environment. The notable contributions of this paper are highlighted as follows,

\begin{itemize}
\item Starting with a review of the main 5G new technologies and target services, existing open challenges in the deployment of 5G networks and the research challenges that may go beyond 5G networks are thoroughly reviewed along with the potential future enabling technologies.
\item The requirements, challenges, and solutions associated with 5G rollout are thoroughly reviewed. In this regard, a comprehensive study on the potential sharing of network infrastructure, public infrastructure, radio spectrum, and generated-data for reducing the 5G deployment cost and developing a sustainable 5G business is conducted.
\item Along with a discussion on the potential barriers in sharing of data in 5G and beyond networks, the state-of-the-art on data privacy and security techniques and their importance in data-sharing based business models is thoroughly reviewed.
\item  State-of-the-art of spectrum trading and management techniques is reviewed, and the analysis is further extended to motivate the location-based shared licensing of the radio spectrum.
\item A framework for passive infrastructure sharing (e.g., public infrastructure, site sharing, mast sharing, power cabling sharing, etc) and neutral hosting is proposed to reduce the deployment cost and provide an opportunity to the local authorities to become direct or indirect partners in the 5G business model.
\item A case study, based on a 5G testbed environment, is conducted for a typical city of UK. The infrastructure sharing potential in reducing the 5G deployment cost in the UK as compared to its anticipated cost is studied. Moreover, the data-sharing and location-based licensing are motivated to bring a significant further reduction in the deployment cost as well as to provide a long-term sustainable business model.
\item Finally, based on the proposed case study and conducted analysis, a data economy based long-term 5G business model is proposed and a list of related recommendations is provided.
\end{itemize}

The rest of the paper is organized as follows. Section II provides an overview of the technologies, services, and open challenges in 5G networks and beyond while Section III presents the deployment requirements, economic constraints, and the potential solutions. Section IV presents a 5G testbed environment learnt prospective for a typical UK city along with a thorough analysis and list of recommendations. Finally, the paper is concluded in Section V.


\section{5G Wireless Networks and Beyond}
 The 5G wireless communication networks have introduced various new revolutionary technologies along with the evolution in the existing networks. The 5G networks are envisaged to offer various new services of new types to everything at all-time with ultra-reliable, ultra-fast, and ultra-low-latency communication links. The standardization of 5G networks as the standalone and non-standalone operating network has appeared with Release 15 \cite{3GPPrel15} of 3GPP named as 5G New Radio (NR). This initial standardization effort is expected to mature with the Release 16 of 3GPP by the year 2020. The test and commercial deployment of 5G NR has now started in some cities of the world. This section provides an overview of the key 5G technologies and services. The challenges associated with the deployment of 5G networks and various other open research challenges (that may go beyond 5G (B5G) networks) are highlighted in this section.


\subsection{5G Target Services}

This section provides an overview of the prime 5G target services, i.e., eMBB, mMTC, and URLLC. Moreover, their enabling technologies and application scenarios are also briefly discussed.

\paragraph{Enhanced Mobile Broadband (eMBB)}

The enhanced mobile broadband (eMBB) in 5G networks targets to provide an increase of $1000\times$ and $10\times$ in aggregate and individual-link throughput \cite{AndrewsJSAC5G}, respectively, compared to the 4G wireless networks. The downlink and uplink data rate targets of 5G networks are up to 20Gbps and 10Gbps, respectively. This high data rate is envisioned to support high throughput demanding services, e.g., tactile internet (TI), augmented reality and HD video streaming. TI is a new 5G service that aims at providing a real-time interface for humans and machines interaction  \cite{Simsek2016tactile}. The interface may support real-time audiovisual and haptic inputs based controlling of machines, e.g., remote humanly controlled robots for industrial and other operations, etc \cite{8452975}. The notable new technologies enabling such high throughput in 5G networks can be named as massive multiple-input multiple-output (mMIMO) and mmWave band.

\paragraph{Ultra Reliable Low Latency Communications (URLLC)}

The 5G wireless networks target at achieving packet error rate and end-to-end latancy of $\leq 10^5$ and ~1ms, respectively. Such ultra reliable low latency communications (URLLC) will help in realizing the dreams of various new types of network services, e.g., auto-driving cars, remote health services (ambulance aid, robotic surgeries etc), logistics automation, to name a few \cite{8467353,IMTvision,8403963}. The critical technology innovations enabling URLLC in 5G networks are network slicing (NS), network softwarization, network function virtualization (NFV), and mobile edge computing (MEC).

\paragraph{Massive Machine Type Communications (mMTC)}

The target of mMTC in 5G networks is the provision of internet access to a massive number of low data-rate and low power devices, e.g., the requisite connectivity to IoT devices. The communication in such typical applications is usually only occasionally required, e.g., in remote environmental sensing and utility metering applications, etc. The key technology enablers for mMTC in 5G networks providing minimal operational cost, grant-free, and time alignment-free connectivity to a massive number of devices can be listed as non-orthogonal multiple access (NOMA), end-to-end (E2E) network slicing (NS), collaborative edge and cloud computing framework, and network function virtualization (NFV) \cite{Bockelmann2018mmtc,Sharmaedgecloudedge}.

\subsection{5G New Radio Technologies}

5G NR is a new radio interface released by 3GPP to satisfy the growing needs of radio access in future wireless networks. The 5G-NR provides a number of significant new technologies and advantages compared to the 4G networks. In the following, we highlight the key features of different evolved and new revolutionary technologies in the 5G NR.


\paragraph{mmWave}

The scarcity of conventional microwave band has led to the exploration of mmWave realm. The multi-gigahertz bandwidth available in mmWave range has a strong potential in addressing the capacity demands of 5G and beyond wireless networks. The initial standardization of mmWave technology for short-range communications initially appeared in IEEE 802.11ad \cite{6736746}.
The Release 15 of 3GPP has specified 24.25GHz -- 52.6GHz band associated with the band numbers from 257 to 511 as one of the major 5G frequency ranges \cite{sanfilippo2018concise}. The propagation behavior of mmWave spectrum in terms of high pathloss and the dominant phenomenon of specular reflections (instead of scattering, as in microwave band) has confined its applications to short-range and line-of-sight (LoS) communications \cite{8594703}. Establishing a better understanding of propagation behavior of mmWave and beyond bands (e.g., sub-teraHz and teraHz bands) may lead to the availability of more usable bandwidth for B5G networks in the future.

\paragraph{Massive-MIMO}

Massive multiple-input multiple-output (mMIMO) \cite{6736761} is defined by a large-scale multi-antenna system in which massive amount of antennas are employed at the base station (BS), i.e., significantly larger than the users being served. The use of large-scale multiple antennas allows the aggressive manipulation of angular/spatial domain. The manipulation of angular domain parameters helps in countering the time and/or frequency selective behaviour of the propagation channels \cite{5462882,mansoor2017massive}. This additional degree-of-freedom (DoF) offered by mMIMO systems can be exploited for diversity, multiplexing, and/or beamforming gains. The 5G NR with mMIMO can exploit 3D beamforming with up to 256 antenna elements at the BS to increase the coverage and capacity of the network \cite{dahlman20185g}.

\paragraph{NOMA}

The provision of wireless medium's access to multiple users was conventionally achieved through the allocation of (sliced) distinct channel resources (time, frequency, or code, etc.) among the users and spatial re-utilization of the resources. The massive growth in the number of network users and the limited availability of usable frequency spectrum has led towards the evolution of multiple-access methods from orthogonal to non-orthogonal resource allocation based methods. The NOMA scheme allocates non-orthogonal channel resources to the users while exploiting an additional dimension/domain of power. Signal processing methods for channel estimation and interference suppression are the prime operations required in NOMA transceivers. Practically, the hybrid of conventional orthogonal multiple access schemes (e.g., CDMA, SDMA, etc.) and power domain multiple access is also referred to as hybrid NOMA. The chipset hardware support for the user side to perform successive interference cancellation (SIC) is also released \cite{8357810}. Along with many other new technologies, NOMA is also one of the technology revolutions being introduced for the first time in 5G \cite{7263349}. The spectrum efficiency and grant-free access provision advantages offered by NOMA makes it one of the core mMTC enablers in 5G wireless networks.

\paragraph{Full-Duplex}

Full-duplex is a key 5G technology, which theoretically has the potential of doubling the channel capacity through the concurrent transmission and reception of information in a single channel resource. The performance of full-duplex method relies on the performance of self-interference-cancellation methods, which may practically be performed through active analog/digital cancellation or passive cancellation methods \cite{8396262}.






\paragraph{Small-Cells}

The ultra-dense network (UDN) planning enables the deployment of multiple small cells within the coverage region of a macro-cell \cite{8516948}. These low-power short-elevation BSs based small cells employ a more aggressive spatial reuse of the resources and efficient users to BS association based on promising propagation channel conditions. The ultra-dense deployment of small cells is a crucial enabler for eMBB in 5G networks \cite{7422408}.

\paragraph{Network Slicing, Network Softwarization, and Mobile Edge Computing}

The new novel concepts of network slicing (NS) and NFV constitute an important part of the list of revolutionary technologies of 5G. In NFV, various network service features are designed as implemented in software that runs on off-the-shelf hardware \cite{7994617}. The examples of these service feature include caching, network address translation, and domain name services. The NS allows the operations on an infrastructure shared among multiple network slices to create an end-to-end (E2E) true virtual network \cite{8320765}. Mobile Edge Computing (MEC) is another exciting technology which offloads the network traffic by serving the users demands directly from the network edge, e.g., this can be achieved through caching of the popular (users specific) contents on the network edge (e.g., BS, access points, etc) \cite{Abbas2018mec}. These technologies hold the key role in achieving URLLC in 5G networks.


\subsection{5G and Beyond Open Challenges}

Despite that 5G networks have introduced various revolutionary technologies and promising services; the huge cost of deployment associated to 5G networks and further growing demands of rigorous network performance requirements necessitate the initiation of dedicated efforts to indicate and resolve the challenges. This section highlights the challenges in the deployment of 5G and some open research challenges and technologies for B5G wireless networks.

\paragraph{Energy Efficiency}

The energy efficiency of communication networks has a direct relationship with the operational, environmental, and economic factors. The 5G technology innovations, such as mMIMO and ultra-dense networks, provide a manifold increase in energy efficiency. A further improvement in network operational energy efficiency in densely connected networks of the future will only help in reducing the network operational cost but also in realizing the dream of a green world.
With the advent of mMTCs and small-cells in 5G networks, a huge number of BSs, sensing nodes, and user devices will constitute the network.
In this massive connectivity context, the energy-efficient design of the network devices will be a highly compelling aspect \cite{7446253}.
The realization of joint energy, spectral, and spatially efficient green communication systems of the future will be in the optimization of b/s/J/m$^3$.


\paragraph{Network Capacity}

With the advent of mMTC in 5G networks, a massive proliferation in the number of network nodes is expected in the coming years. The densification of cells may eventually meet its practical limit, as a further decrease in cell size (towards very tiny cells) may have some unmanageable associated physical deployment constraints, deployment costs, and inter-cell interference. Further evolution and revolution in network technologies and extension in usable frequency spectrum may be required. A promising future enhancement can be seen in the idea of employing vehicular BSs, i.e., cell-free networks served through Unmanned Aerial Vehicles (UAVs). The future networks may be 3-D natured with volumetric quantification of coverage specifications and spectral efficiency \cite{QML_6G_Junaid} (i.e., b/s/Hz/m\textsuperscript{3}).

\paragraph{Throughput}

The throughput offerings of 5G wireless networks are expected to attract various high data rate demanding applications, e.g., virtual reality applications. Such applications and massive connectivity may result in 5G reaching its limit in a decade or so \cite{8760275}. For meeting the exceptionally high throughput demands of future networks, one venue can be the exploration of mmWave, sub-teraHz, and teraHz bands. Moreover, communication in the optical spectrum (visible light) may also attract various LoS and indoor communication applications \cite{8782879}.

\paragraph{Inter-portability and Congestion in Heterogeneous Networks}

The 5G networks will operate in the coexistence of its predecessors (1G to 4G).
Achieving harmonization of operations across different network architectures with different conditions for real-time communication applications is another challenging issue for operations in 1G-5G hybrid heterogeneous networks.
The MNOs may deploy a core 5G radio access network (RAN) infrastructure to coordinate the network heterogeneity \cite{obiodu20175g}. The application of machine Learning (ML) methods for achieving the harmonization and learning of the network state is a potential enabler \cite{QML_6G_Junaid}.

The MEC technology of 5G wireless networks is expected to provide a substantial improvement in the performance of not only to 5G users/devices but also to users operating on previous generations \cite{Abbas2018mec,QML_6G_Junaid}. The intelligent caching at the network edge will help in offloading the data traffic from the network backhaul. The congestion in access networks in ultra-dense connectivity scenarios (e.g., mMTC, etc.) is another challenging issue. Grant-free access to IoT devices may help in resolving the congestion in ultra-dense access networks imposed due to the heavy burden of signaling for enabling communication of a large number of short-data-packets.

\paragraph{Ethics for Big Data Analytics}

The recent advances in communication network technologies, proliferation in the number of connected devices, and growing multimedia applications are leading towards a flourishing expansion in the data generation \cite{8304385}. The radio communication networks are not only the carriers but also a leading source of generation of data. Appropriate exploitation of big data analytics has a strong potential in facilitating the improvement in the performance of the communication systems as well as in maximizing the revenue generation opportunities for the stakeholders \cite{7429688}.
In \cite{Sharmaedgecloudedge}, the data-aware intelligence for extracting useful information from the data and enhancing the network performance for IoT applications have been discussed.  Also, a collaborative processing framework while combining the benefits of edge and cloud computing for live data anaytics in IoT networks has been proposed.
Along with the benefits offered by big data analytics, there are also various critical concerns being raised regarding the ethics of the analytics \cite{7515114}. The essential factors for devising comprehensive data sharing policies need to be explored to interpret the broader context of data choice, collection circumstances, ownership rights, substantiation and usage permissions. There is a need to conduct thorough investigations to understand the implications related to data analytics technology in B5G communication networks concerning the individuals and organizational interests.

\paragraph{Security and Privacy}

The provision of security and privacy of data is among the most important concerns in 5G rollout. In densely connected networks, provided the provisions of data security and privacy, the enormous amount of generated data can be potentially used to enhance the network performance \cite{8428412} as well as revolutionize the existing business models.
Devising a data value generation based long-term business model for 5G may help in attracting the operators to invest the requisite revenue for 5G rollout. However, ensuring a balance in the policies for necessary data sharing with high data security may emerge as a main policing challenge in the future. A thorough review on security and privacy related challenges in 5G and beyond networks is provided in the sequel.

\paragraph{High Deployment cost and Lack of Business Models}

The huge CapEx and operational expenditure (OpEx) requirements and lack of clear business model are among the major challenges in fully benefiting from the proposed 5G technologies. The prominent directions to reduce the deployment costs can be stated as infrastructure (passive and active) sharing, neutral hosting, and location-based spectrum licensing. The deployment cost analysis and proposals for the solutions are one of the primary focus of this papers, the details are discussed in the following dedicated sections.



\section{5G Deployment Requirements, Challenges, and Solutions}

This section mainly discusses the 5G deployment requirements, the economic constraints associated with the 5G rollout, and their potential solutions.
The principal factors that may affect the speed of adoption of 5G and its generation of value can be listed as the cost of 5G rollout, availability of opportunities for different 5G service classes, need for a supportive policy framework, availability of 5G devices, compelling business model and value perception \cite{Report_MobileEconomy2019}.
We propose that smart management and sharing of infrastructure (network and public), generated data, and radio spectrum between different users/operators are the potential directions to reduce the 5G deployment cost as well as develop long-term 5G business models. To this end, this section first thoroughly investigates the potential in sharing of network and public infrastructure for reducing the 5G rollout cost. Furthermore, the potential of smart management and sharing of the radio spectrum in reducing the 5G rollout cost is reviewed. The sharing of data to generate value as well as improve network performance requires comprehensive data privacy and security model. Therefore, security and privacy concerns and other barriers in sharing of data are also thoroughly studied in this section.

\subsection{Infrastructure Requirements}

Over the past few decades, the land-mobile radio cellular networks have evolved from 1G to 4G. The infrastructure for these mobile generations has been established with a massive investment. In the existing infrastructure, there may exist over several hundred thousands of wireless communication sites worldwide. The sites may include wireless setups for emergency services, broadcast services,  cellular communication services, and other national communication services. The reduction in a typical cell-size has been witnessed along with the progression in generations of wireless networks. The cell coverage radius of a standard cell from 2G to 4G has evolved from approximately 10s of km to a single km. To fully harvest the benefits from 5G technologies, an extension in the spectrum and densification of the network is obliged. The inclusion of radically high mmWave radio spectrum in 5G requires beamforming and beam-management equipment at the small-cell BSs.
Furthermore, the ultra-dense deployment of cells, i.e., small-cells with cell-radius of only a few meters, necessitates massive infrastructural provisions. Consequently, to meet the vast infrastructural requirements for deploying small-cells and mmWave communication facilities at the ultra-densified BSs in 5G networks, an enormous revenue investment is needed.

\begin{table}[t]
  \centering
  \caption{Safe work distance recommended by ICNIRP from the typical transmit power of the respective frequency \cite{ICNIRP_guideline_distance_98}}
    \begin{tabular}{|c|c|c|c|}
    \hline
    \multicolumn{1}{|p{6.145em}|}{\textbf{Frequency (MHz)}} &
      \multicolumn{1}{p{6.73em}|}{\textbf{ICNIRP EMF Limit (V/m)}} &
      \multicolumn{1}{p{4.985em}|}{\textbf{EIRP (W)}} &
      \multicolumn{1}{p{5.715em}|}{\textbf{Distance (cm)}}
      \bigstrut\\
    \hline\hline
    \textbf{800} &
      39 &
      100 &
      140.44
      \bigstrut\\
    \hline
    \textbf{900} &
      41 &
      100 &
      133.59
      \bigstrut\\
    \hline
    \textbf{1800} &
      58 &
      50 &
      66.78
      \bigstrut\\
    \hline
    \textbf{2100} &
      61 &
      50 &
      63.49
      \bigstrut\\
    \hline
    \textbf{2600} &
      61 &
      10 &
      28.39
      \bigstrut\\
    \hline
    \textbf{3500} &
      61 &
      2 &
      12.7
      \bigstrut\\
    \hline
    \end{tabular}%
  \label{tab_safe_distance}%
\vspace{5pt}
  \centering
  \caption{Typical mapping of installation classes for typical small cell deployments \cite{SCF_19}}
    \begin{tabular}{|c|p{3.145em}|p{3.8em}|p{4.0em}|p{3.0em}|p{3em}|}
    \hline
    \multicolumn{1}{|p{3.5em}|}{\textbf{3GPP BS Class}} &
      \textbf{Config-uration} &
      \textbf{Typical Total Tx Power} &
      \textbf{Typical Gain} &
      \textbf{EIRP Range} &
      \textbf{Insta-llation Class} \bigstrut\\
    \hline \hline
    \multicolumn{1}{|c|}{\multirow{2}[4]{2.3em}{Medium Range BS}} &
      2 bands &
      20 W &
      7 - 13 dBi &
      100 - 400 W &
      E+
      \bigstrut\\
\cline{2-6}     &
      1 band &
      10 W &
      7 - 13 dBi &
      50 - 200 W &
      E100 or E+
      \bigstrut\\
    \hline
    \multicolumn{1}{|c|}{\multirow{2}[4]{2em}{Local Area BS}} &
      5 bands &
      2.5 W &
      2 - 5 dBi &
      4 - 8 W &
      E10
      \bigstrut\\
\cline{2-6}     &
      1 band &
      0.5 W &
      2 - 5 dBi &
      0.8 - 1.6 W &
      E0 or E2
      \bigstrut\\
    \hline
    \multicolumn{1}{|c|}{\multirow{2}[4]{2em}{Home BS}} &
      5 bands &
      100 mW &
      0 - 3 dBi &
      0.1 - 0.2 W &
      E0 or E2
      \bigstrut\\
\cline{2-6}     &
      1 band &
      20 mW &
      0 - 3 dBi &
      0.02 - 0.04 W &
      E0
      \bigstrut\\
    \hline
    \end{tabular}%
  \label{table_small_cell_classes}%
\end{table}%

In the provision of true 5G services, an ultra-densification is inevitable, and meeting the challenges in providing essential infrastructural requirements needs the conduction of thorough investigations. In such an ultra-dense BSs deployment context, the conventional BSs deployment-related challenges may also get amplified. These challenges include: abidance of BSs antenna tilts and power radiation standards for health/safety, lease disputes with landlords, planning permissions and lack of suitable sites. The safe user to BSs distance can be calculated by exercising the recommendations for electric field intensity by the International Commission on Non-Ionizing Radiation Protection (ICNIRP) \cite{ICNIRP_guideline_distance_98}. For radiations at different frequency bands, Table \ref{tab_safe_distance} presents the electric field intensity (using ICNIRP recommendations), equivalent isotropically radiated power (EIRP), and safe-user-distance in volts per meter (V/m), Watts (W), and centimeter (cm), respectively. The basic deployment principle being followed by the operators is to ensure 1/10\textsuperscript{th} of the ICNIRP voltage density, for the exposure of RF for a long time (i.e., more than six minutes). Thus, considering the radiation power, the safe-user-distance in the table represents the shortest permissible distance between the transmitter and human-head. The enforcement of maximum power radiation standards and safe-user-distance assurance for health/safety may be another critical challenge in ultra-dense deployment context. The details of different classes of BSs defined by 3GPP that can be manifested as small-cells for ultra-network-densification, as provided by \cite{SCF_19}, are presented in Table \ref{table_small_cell_classes}. The typical BS bands, transmit power, gain, EIRP range, and installation classes configurations are classified into medium range, local area, and home BSs. The details provided in Table \ref{tab_safe_distance} and \ref{table_small_cell_classes} may assist in devising a comprehensive strategy for infrastructure sharing and value generation opportunities to support a swift 5G rollout.

\begin{figure}
  \centering
  \includegraphics[width=\columnwidth]{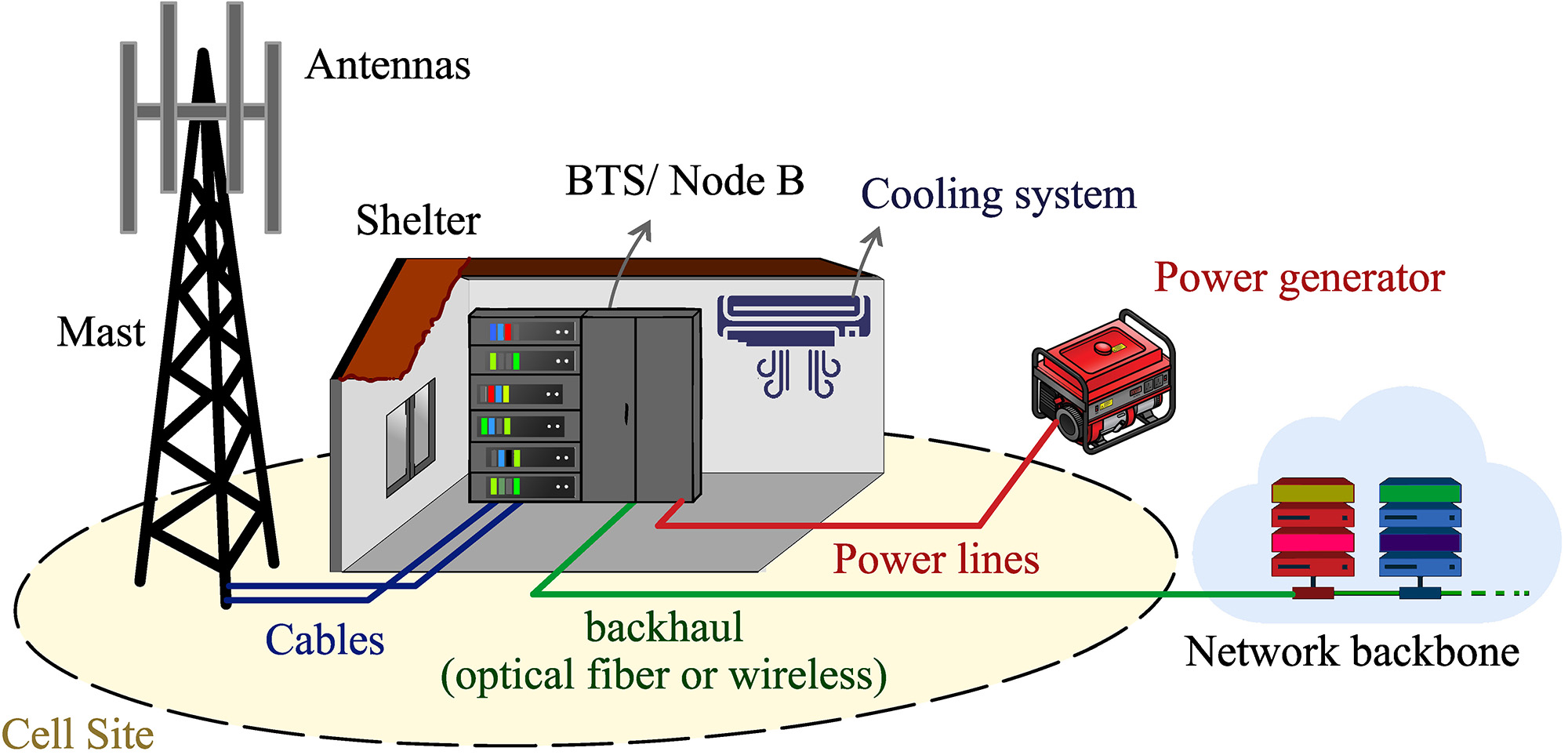}
  \caption{Schematic diagram of a cell site illustrating the components in the context of possible infrastructure sharing.}
  \label{fig_schematic_s9}
\end{figure}

\begin{figure*}[th]
  \centering
  \includegraphics[width=\textwidth]{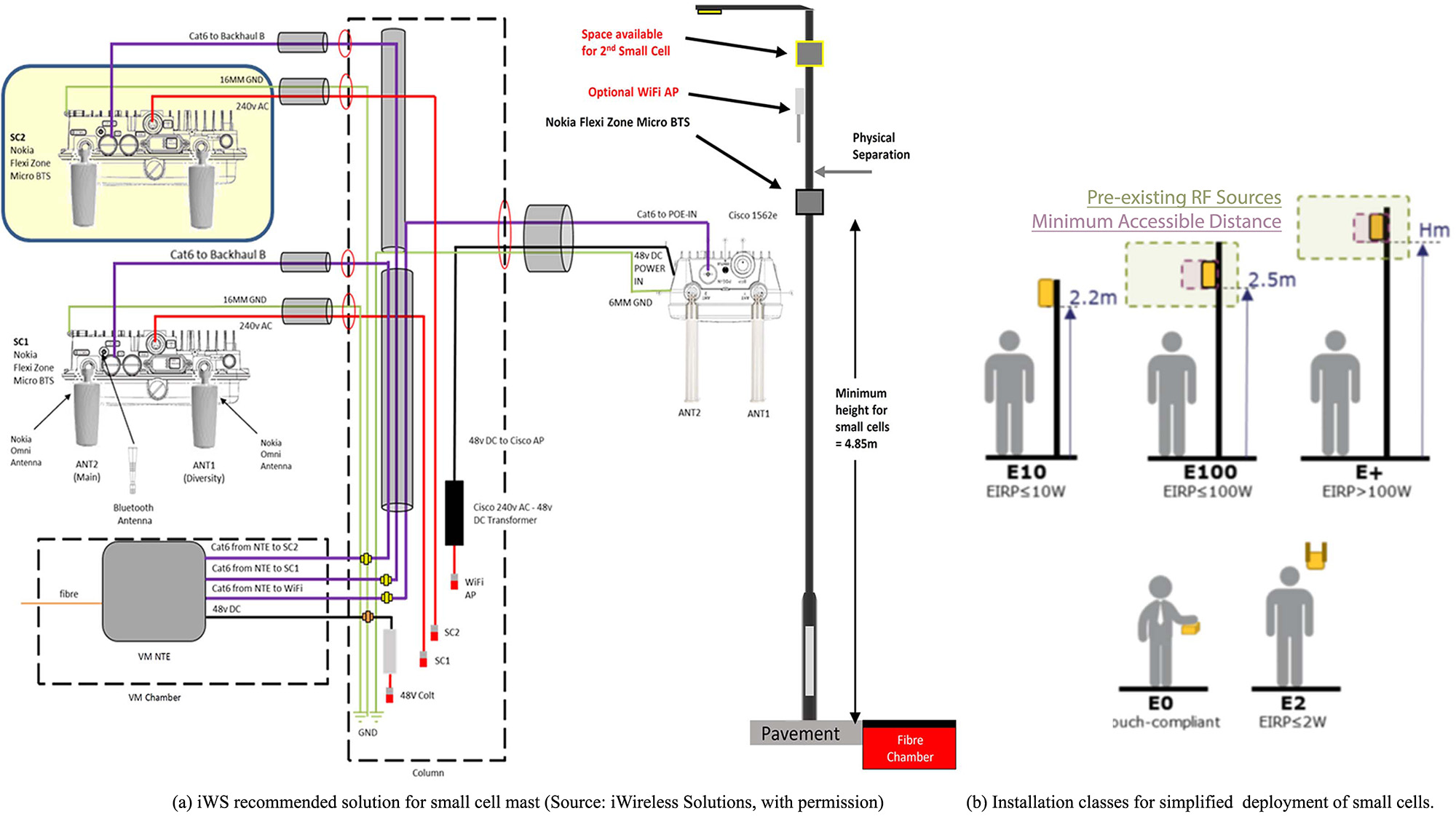}
  \caption{iWS recommended solution for small cell mast \cite{iws_sol_1} and installation classes for simplified deployment of small cells.}
  \label{fig_poll_s8}
\end{figure*}

\subsection{Infrastructure Sharing Potential}

Sharing of existing telecommunication and public infrastructure for small-cell deployment is an attractive solution to substantially reduce the 5G deployment cost and to encourage revenue generation opportunities for Mobile Network Operators (MNOs).
The hesitance of MNOs in implementing the standalone 5G solution is understandable.
Fig. \ref{fig_schematic_s9} illustrates the schematic diagram of network infrastructure by highlighting the important candidate components for possible infrastructure sharing.
To this end, the solution may be in the optimization of technology and passive infrastructure along with the development of the revenue generation models for supporting the sharing of active infrastructure. To make the cost-optimization more effective and quick, devising a robust pathway is vital. In this regard, the promotion of competition among different partakers from infrastructure domain may help. Moreover, the thorough policying for sharing infrastructural services between different MNOs is another potential step. More importantly, the induction of non-MNO assets for 5G deployment into the infrastructure sharing business model will help in further stimulating the rollout. The Neutral Hosting model can be mainly signified as an effective form of infrastructure sharing.

\emph{Active sharing} requires the MNOs to share elements of the active network layer. The active sharing components include RF antennas, MSC, HLR, OMC, SGSN/GGSN, core transmission ring, core network logical entities, billing platform and value-added services (VAS). Despite the cost optimization advantages associated with active sharing approach, it is becoming unpopular due to various dynamic reasons linked to network infrastructure domain. The forms of \emph{passive sharing} are site and mast sharing. The passive sharing agreements are associated with multi-tenancy within sites (physical location), power cabling, cabinet/ shelter, generator, cooling system, and mast and backhaul (fiber).

The following is the potential list of sharing components.
\begin{enumerate}
  \item Mast Sharing
  \item Site Sharing
  \item Full RAN Sharing
  \item Network Roaming
  \item Core Transmission Ring Sharing
  \item Shared Core Network Elements and Platforms
\end{enumerate}

Moreover, the public infrastructure from local governments potentially available for sharing/reuse can be listed as:

\begin{enumerate}
  \item Streetlamps
  \item Road/street signs
  \item Rooftops
  \item Tall building with suitable projections surfaces –- considering proximity
  \item Traffic signals
  \item Cabinets
  \item CCTV installations
\end{enumerate}

For sharing the indicated public infrastructure, there are various other factors which may need careful consideration. The capability of street furniture (e.g., streetlamps) for bearing the extra weight required to hold antennas and other equipment needs to be assessed. Moreover, the height requirements of the street furniture to facilitate different BS installation classes to meet the safe-user-distance recommendations (as discussed in the previous subsection) also needs to be investigated.
Furthermore, various other factors like weather aspects (e.g., aerodynamics etc.) are also critical in adopting a passive infrastructure sharing option. In light of these critical factors, streetlamps are recommended as the most suitable choice for passive infrastructure sharing in this work. The iWS recommended design \cite{iws_sol_1} for a typical small-cell mast is illustrated in Fig. \ref{fig_poll_s8} along with the height recommendations for different BS installation classes. This solution is for 4G cells, which can be revisited in the context of required ultra-densification in 5G networks.

\subsection{Data Sharing Potential and Associated Barriers}

A significant increase in the generation of data from different network services is expected in the coming years \cite{QML_6G_Junaid}. An annual growth of 55\% in data-traffic is forecasted from the year 2020 to 2030 \cite{union2015imt}, which will include a significant contribution from the subscriptions of new services introduced by 5G such as mMTC. As a result, the data generated per month is expected to reach $5.016$ ZetaBytes (ZB) by the year 2030. Appropriate exploitation of this huge amount of generated data can assist in not only improving the user experience but also in creating the new revenue generation opportunities. A balanced policy for sharing of data with different stakeholders, while also maintaining the necessary privacy and security provisions, can help in directly translating the data value into performance enhancement and revenue generation opportunities. To this end, this section highlights the barriers in data sharing and motivates the opportunities for their resolution.

\paragraph{Lack of use cases} There are no explicit motivational use cases available; which is because of the data value conception being a recently emerged perspective. This data-value based core business has developed with a drastic increase in the amount of data generation arose with the growing network services. Development of explicit examples/models for facilitating successful selling and buying of valuable data in the mutual benefit of all stakeholders is necessarily required.

\paragraph{Immature market}
The current business strategy is not exclusively based on the data value. The immaturity of the market in this context is one of the critical barriers in the initiation of dedicated efforts for defining the required policies of data sharing. The commissioning of data sharing has a strong potential in generating new data value based business models in the long-term.

\paragraph{Fragmented data landscape}
The data in the current model is often not in the entangled form, which is a primary requirement to identify the data value. The lack of availability of concatenated/de-fragmented data to define a meaningful business model is a crucial barrier.

\paragraph{Reluctance of data sharing}
The data privacy, security of data privacy, and the absence of any comprehensive business model are among the primitive reasons causing hesitation in the sharing of data. The lack of a distinct definition of direct or indirect financial benefits associated with data sharing for core business is another leading cause.

\paragraph{Skills and Competence}
The lack of skill sets for the collection and processing of data into suitable compositions for enabling data sharing is another critical barrier. The shortage of available data scientists and the high cost of associated equipment are vital causes. Moreover, the absence of a unified technical-platform for data processing and serving is also a prime contributor to the barrier.

\subsection{Security Challenges}
The provision of security and privacy in massively connected 5G and B5G communication networks is one of the biggest challenges of the future. 5G offers new and disruptive business cases \cite{8648405}. While rolling out 5G networks in the next-generation smart cities in general and UK in particular, due to many new features of 5G networks and the fact that 5G will envelop almost all dimensions of human life, business, and government affairs with ultra-high-speed access to services, anywhere, anytime and any type, 5G security requirements have to be thoroughly researched \cite{7345407}. 5G rollout will also bring a wide range of threats and a greatly expanded attack surface \cite{8334918}. For example, 5G physical layer security at the RAN \cite{7345407} is susceptible to new types of attacks and performance bottlenecks such as eavesdropping, contaminating, spoofing, and jamming. Since 5G rollout will leverage existing telecom infrastructure and other computing and networking paradigms \cite{7345407}, 5G will be live within heterogeneous networks (HetNets) \cite{7081072}.

The HetNet architecture, compared to a single-tier architecture, can potentially lead to the UEs to be more vulnerable to eavesdropping \cite{8125684} and hence privacy and location leakage may arise due to frequent handover caused by the high density of small cells in the HetNet \cite{7081072}. As a result, a new generation of security services has to be offered \cite{8125684}. Novel 5G AKA, USIM and ECC based design of handoff authentication for 5G-WLAN HetNets will be needed that can extend the provisions of secure and seamless internet connectivity \cite{8706883}. Many of these additional requirements come from the technology shift to SDN \cite{7345407} and NF virtualization (NFV) \cite{7994617}, network slicing, massive MIMO \cite{6736761}, NOMA \cite{7842433}, ultra-dense small cell network \cite{8516948}, D2D and M2M communications, and the cloud, and they lead to the need of increased security on the network side. 5G NOMA, mmWave, massive MIMO, and beamforming can improve physical layer security of 5G networks through co-operative jamming \cite{8125684,8792139}, which will allow secret and high-quality channel with the legitimate UEs while frustrating eavesdroppers with noisy, random, and poor channel conditions. The directional property of mmWave can be leveraged to establish and share secret keys that are unconditionally secure from the passive eavesdroppers\cite{8594703,8684919}.

A variety of emerging new use-cases and networking paradigms demand new security requirements and considerations  \cite{8125684}. For example, 5G networks need to employ adaptive intrusion detection system, which can perform the following tasks: (i) detect bandwidth spoofing attack on 5G relay, small cell access point, and base station, (ii) employ UEs initial authentication at the access points, and 5G RAN by a highly secured authentication and handover mechanism with the minimal handover latency and no loss of user privacy. The SDN controller handles DDoS attack and a secure VNF in the cloud filters out malicious packets from legitimate packets \cite{8523804}. 5G device-to-device (D2D) and machine-to-machine (M2M) communication security needs, vulnerabilities, and potential solutions have to be investigated. To this end, the important contexts include the following: (i) direct radio communications, (ii) large-scale D2D deployments, (iii) cooperative communications for securing D2D communications, (iv) power control and channel access in securing D2D communications,(v)  continuous authenticity with legitimacy patterns, (vi) key exchange protocols involved with the D2D users and gNB, (vii)  design of D2D links to use as friendly jammers, and (viii) helping authorized cellular users against malicious wiretapping \cite{8125684,8335294}.

Furthermore, C-RAN security has to be ensured in service plane, control plane, and physical plane \cite{7954591}. 5G NR vulnerability has to be addressed in terms of jamming and spoofing by investigating the physical downlink and uplink control channels/signals and through designing proper mitigation strategies for proceeding towards the design of 5G NR chipsets and BSs \cite{8403769}. New 5G paradigms will have to be handled by the UK MNO businesses. For example, security challenges related to network slicing; such as on-demand security isolation of network slices, the security of inter-slice communications, impersonation attack, security policy mismatch, DoS attack, side-channel attacks, privacy attacks, resource harmonization between inter-domain slice segments and hypervisor attacks have to be considered \cite{8334918,8792139}. New types of 5G-based verticals, e.g., IoT, need to establish secure 5G-based network slicing technique with secure key establishment among IoT devices, MEC, and IoT cloud server \cite{8314666}. UK businesses will need innovative 5G Network Slice brokering mechanism using blockchain for reducing service creation time and for enabling the manufacturing equipment to autonomously and dynamically acquire the slice required for more efficient operations \cite{8260929}.

\begin{figure}
  \centering
  \includegraphics[width=\columnwidth]{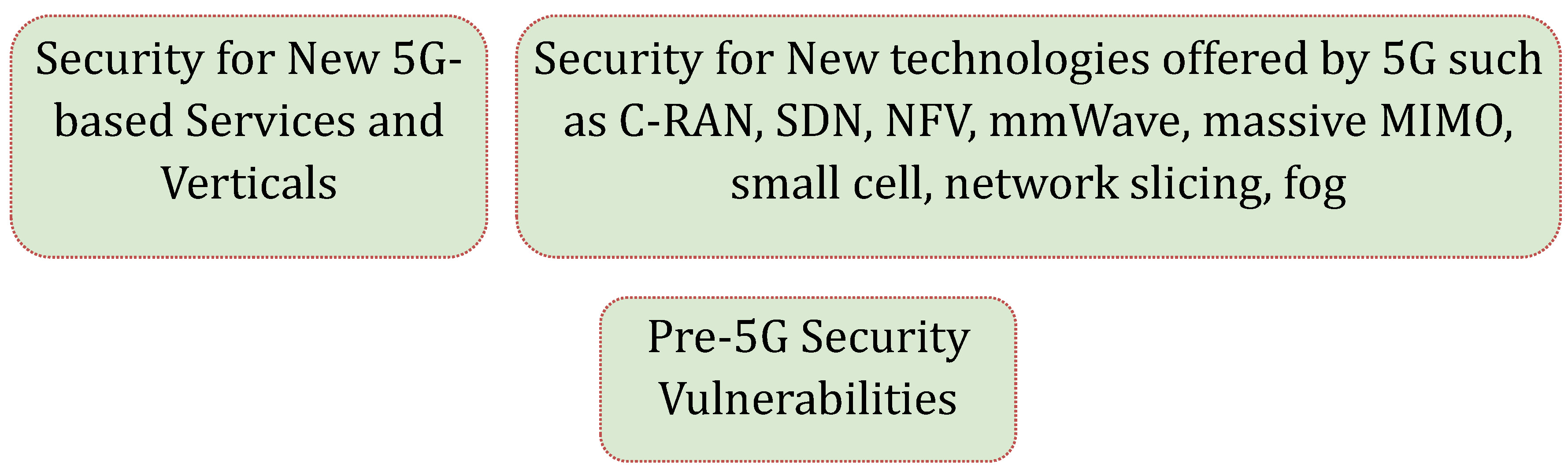}
  \caption{5G and Beyond Roll out Security Consideration Scope}
  \label{fig_security_1}
\end{figure}

Moreover, SDN and NFVs are vulnerable to various types of attacks including the following: (i) DoS, hijacking, and saturation attack on SDN controller, switches, and hypervisor, (ii) network slice theft from hypervisor and shared cloud resources and SDN (virtual) switches, (iii) routers configuration attacks, (iv) SDN configuration attacks, (v) penetration attack on SDN virtual resources, (vi) TCP-level attack on SDN controller-switch communication, and (vii) man-in-the-middle attack on the SDN controller-communication \cite{8334918,8125684,8792139,8712553}.

The advent in security mechanisms is highly required to meet the overall 5G advanced features such as ultra-low latency and ultra-high energy efficiency \cite{8125684}. Different to the conventional radio cellular networks, the emerging 5G wireless networks will be service-oriented, which necessitates a particular emphasis on security and privacy requirements from the perspective of service-based architecture (SBA) \cite{8125684}. Hence, UK MNOs have to ensure data exchange security for network function (NF) service registration and de-registration, NF service discovery, NF service authorization and authentication in the presence of different attack vectors that may cause loss of NF availability, loss of data integrity, and attack from the insiders \cite{8771320}. 5G advocates the use of MEC, which offers heterogeneous network nodes inter-operating in an open ecosystem where distributed computing and virtualization may be exploited by service providers to extend the provisions of different applications to the end-users \cite{8334918}. Hence, at the time of rolling out 5G networks, MEC threats have to be considered at the frontend (UE/IoT to MEC network), backend (MEC network to cloud), and 5G MEC core network \cite{8334918,8314666}.

New dimensions of 5G UE Security, trust, authentication, policy, compliance, and privacy for ultra-mobility will be required. Since 5G networks are expected to extend the provisions of everything as a service, where the users data/information will be stored and shared online, maintaining users/data privacy during 5G rollout in HetNets will be highly challenging \cite{8334918}. Due to the reduced cell size in 5G networks, there may often appear the scenarios in which a mobile user may move through multiple small sized cells within a single communication session. Therefore, the privacy assurance is more challenging in 5G networks due to the possible involvement of untrusted or compromised APs involved during the handover \cite{7081072}. Hence, hybrid and flexible authentication of UEs should be devised that will allow authentication by the network only, authentication by the service provider only, and authentication by both the network and service providers \cite{8125684}. 5G rollout cost model in terms of financial protection against secrecy outage and service outage can be considered by introducing a cyber-insurance framework for 5G cellular networks \cite{8335293}. Pre-5G or legacy privacy vulnerabilities should be carefully addressed while rolling out 5G network \cite{8712553}. Pre-5G legacy authentication, privacy, and protocol exploits in the context of 5G have to be addressed such as implicit trust in pre-authentication messages and legacy symmetric key security architecture \cite{jover2019current}. Furthermore, identify security, location security, IMSI security, and mobile terminal security have to be considered \cite{8334918}.

\begin{figure}
  \centering
  \includegraphics[width=\columnwidth]{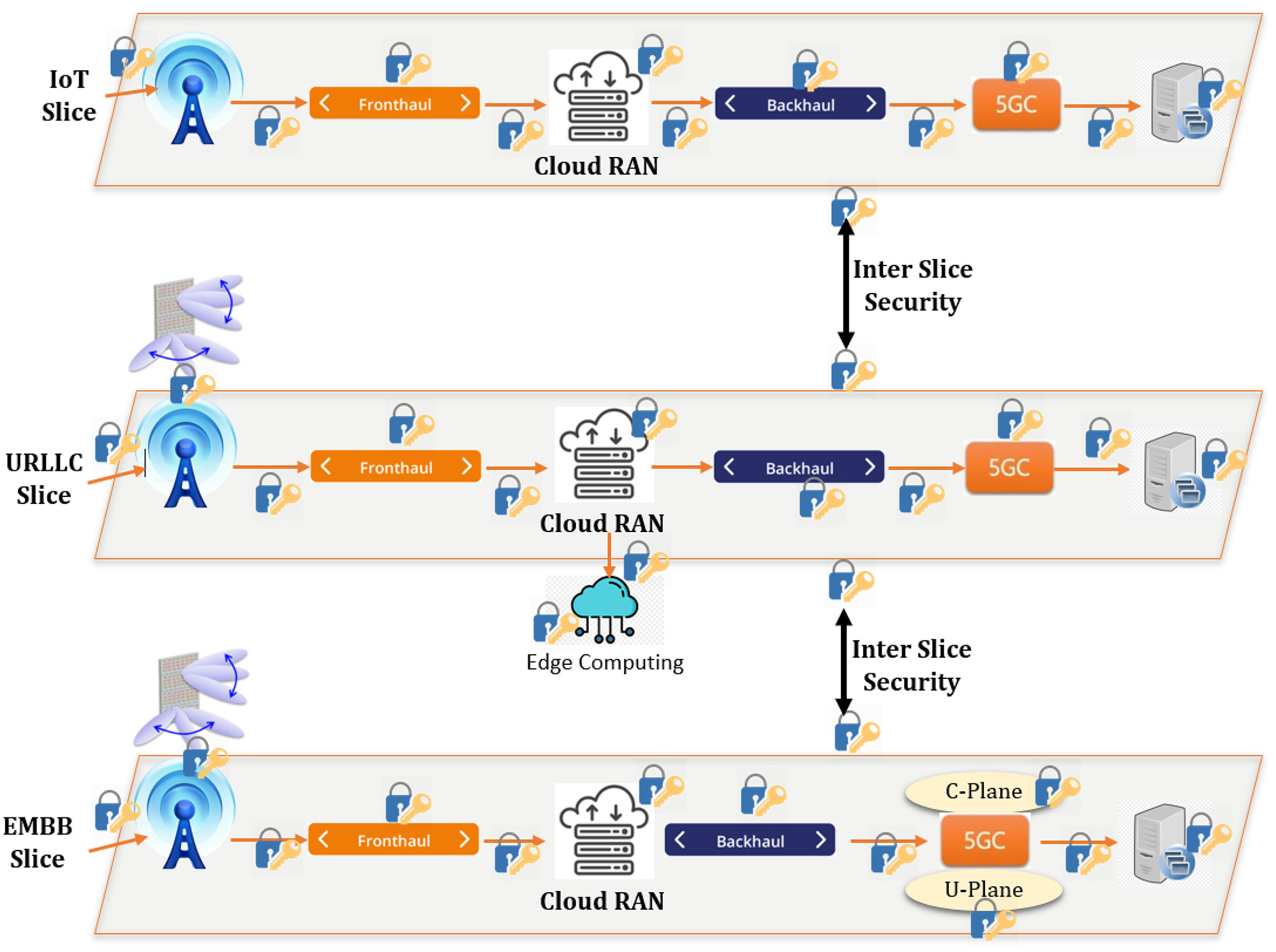}
  \caption{Proposed Security endpoints to be considered while rolling out 5G Infrastructure in UK}
  \label{fig_security_2}
\end{figure}

While rolling out 5G in the UK, the threats originated from devices or UEs, the air interface at RAN threats \cite{8454666}, edge network threats, backhaul threats, 5G core network threats, and external network threats have to be considered \cite{8334918} \cite{8792139}. Because failing to handle the security and privacy of user data and application would have the repercussion of 5G rollout costs. For example, the 5G network has to proactively handle UE threats such as bots, DDos, man-in-the-middle attacks, firmware hacks, device tampering, and malware. The 5G RAN air interface threats include physical layer security to handle jamming, man-in-the-middle attacks, and eavesdropping. The threats at the MEC layer have to handle MEC server vulnerability, rogue nodes, weak authentication issues, side-channel attacks, and improper access control. The backhaul threats include DDoS attacks, control and user plane sniffing, MEC backhaul sniffing, and flow modification attacks. The 5G core network threats that need to be mitigated and handled are software and SDN issues, API vulnerabilities, network slicing issues, DoS and DDoS attacks, improper access control, and virtualization issues. Finally, the external network threats include application server vulnerabilities, cloud services vulnerabilities, bots, and other IP based attacks, application vulnerabilities, API vulnerabilities, and roaming partner vulnerabilities \cite{8792139}.

Different stakeholders related to 5G rollout are recommended to consider the following B5G security considerations to minimize the costs related to security breaches and safety and privacy of data flowing through 5G networks \cite{8712553}. Fig. \ref{fig_security_1} shows a high-level requirement of 5G security considerations, and Fig. \ref{fig_security_2} shows the salient areas of 5G security that needs to be considered while rolling out 5G. In this section, we will illustrate the state-of-the-art 5G rollout security dimensions that need to be considered by the concerned entities.

\paragraph{Blockchain usage for B5G Applications}

UK businesses can use blockchain for a 5G-based decentralized business model such as commerce, context, content, and connection \cite{8648405,8712553}. For example, blockchain smart contract can be used for secure 5G network slice brokering and maintaining immutable service level agreement (SLA) ledgers to bind 5G business actors such as manufacturing equipment owner, IoT manufacturing equipment, infrastructure providers (InP), MNOs, micro-operators ($\mu$O), virtual mobile network operators (MVNO), over-the-top service providers (OTT) and verticals.

Blockchain can be used for secure 5G network sharing scenarios \cite{8685867}, e.g., multiple 5G network operators can extend and collaborate among roaming end-users and incentivizes local businesses and other actors to densify and extend the 5G coverage. Blockchain ledger can be used to store the proof of bandwidth and other types of 5G network resource usage, traffic flow and accounting. MNOs can employ blockchain to provide secure authentication scheme for 5G Ultra Dense Access Point Network \cite{8470085}. Also, blockchain is envisioned to provide security and privacy of IoT data in 5G HetNets \cite{8731639}. In the literature, researchers have proposed blockchain to offer efficient privacy-preserving and data sharing schemes for 5G verticals \cite{8320551}.

UK businesses can leverage blockchain for secure registration of a new cellular user and UE, authentication and authorization of users and different services, usage of networks, distributed mobility management (DMM), authenticate the priorities and 5G network services usage and propose algorithm of allocating communication and computation resources to minimize the delay of data transmission and computation, billing and payment and manage roaming bills context of 5G HetNets \cite{8356363}. Blockchain can handle robust and universal seamless handover authentication for 5G HetNets by leveraging the trapdoor collision property and the tamper-resistant property. More specifically, blockchain will allow 5G network slice providers to securely perform brokering process and allow leasing resources from different providers securely and privately \cite{8707070}. Furthermore, blockchain will allow UK's businesses to enable industry 4.0 automation processes and manufacturing IoT equipment to autonomously and dynamically acquire the 5G-based slices with QoS needed for most efficient operations \cite{8368983}. Blockchain-based authentication and key agreement protocol can be employed, which will allow UEs to move smoothly in a trusted APs group without frequent authentication within an ultra-dense small cell network \cite{8470085}.

\paragraph{Artificial Intelligence, Deep Learning and Machine Learning for 5G Security Threat Intelligence}

Since the UK is one of the leaders in artificial intelligence research, 5G security can leverage this strength. Various use case scenarios of AI-based security for 5G and beyond applications can be found at \cite{8712553}. Blockchain and AI methods can together form a strong platform to support secure and intelligent resource management, flexible networking, and reliable orchestration in 5G and beyond scenarios such as spectrum sharing, D2D caching, V2V energy optimization, and computation off-loading \cite{8726067}. The novel machine learning algorithm can be trained to teach security threats faced by 5G network-interfaced intrusion detection and prevention system, cyber threat and anomaly identification system, and help to secure threats on UK's 5G network \cite{8304873}. UK's 5G MNOs or service brokers can use AI to allow self-adaptation of security needs according to live 5G traffic flowing through VNF. The AI-based VNF can employ auto-scaling and deploy more 5G network resources, employ appropriate deep learning framework, or even the detection model, with a more suitable one to the given cyber-defense context \cite{8283694}.

\paragraph{Quantum Safety and Next Generation Ciphers for B5G Applications}

In order to face the challenge of cryptographic vulnerability threats due to advancements in computing capabilities of adversaries, UK 5G rollout should consider quantum-resistant authentication and data distribution scheme, and lattice-based homomorphic encryption technology, which greatly reduces the network burden at the same time achieves strong security including privacy protection and anti-quantum attacks \cite{8781929}. 5G MNOs can leverage the visible spectrum as a noise source for designing next-generation random cryptographic seeds and key generation system suitable for 5G networks \cite{8276259}. To support real-time data secrecy over 5G intra-slice security applications and protect the private information and hide the communication signals in the frequency, spectrum stream cipher can be a viable option \cite{8315003}. To lower the authentication delay in an ultra-dense small cell network, certificate authority (CA)-based approach can be availed \cite{8718355}. Different 5G network slicing may deploy different public-key cryptosystems, and hence, the 5G network should allow diversified types of cryptosystems that will allow heterogeneous sign-cryption schemes such as public key infrastructure and the certificate-less public key cryptography environment \cite{8268052}.

\paragraph{5G Verticals' Security Considerations}

In general, an integrated effort has to be given to secure the UK's business verticals. In the following, we focus security considerations of rolling out 5G-based V2X and Industry 4.0 verticals.

Security vulnerabilities in the areas of mutual authentication and authorization, confidentiality and integrity protection, replay protection, Secure provisioning and storage, privacy ID, personal data, and tracking in the vehicular context are of significant importance \cite{8088619}. Especially, the security analysis of V2X verticals in the areas of the termination of user plane security at gNB, authentication and authorization of UE at the 5G RAN, 5G RAN security, UE Security, and network Slicing security have to be ensured \cite{8088619}. V2X security can be obtained by securing 5G network slicing through permissioned consortium blockchain. Using dedicated networking slicing and blockchain ledgers, vehicles can share information via 5G networks with outside world entities or D2D entities \cite{8343866}.

A 5G enabled vehicular network can facilitate a reliable, secure and privacy-aware real-time video reporting service by using novel block cipher with 1.2 Gbps speed of secure video data sharing. This is to enable the participating vehicles to instantly report the high-definition videos/photos of any events (e.g., traffic accidents, etc) to ensure a timely response from concerned departments (e.g., sending ambulance vehicles to the accident scene, etc) \cite{7433471}. Using blockchain, SDN-enabled 5G-V2X can detect malicious vehicular nodes or messages while keeping the overhead and impact on the network performance in an acceptable range \cite{8701642}. Leveraging 5G SDN \cite{8792139} with resilient V2X security design, different types of attacks such as DDoS targeting either the controllers or the vehicles in the network can be mitigated, and at the same time, it allows tracing back the source of the attack \cite{7939143}. By leveraging the directional beamforming, secure 5G V2X applications such as vehicle platooning will allow platoons to establish shared secret keys \@ 166 Mbps, which is four-times higher than that of Diffie-Hellman and assumed to allow One Time Pad (OTP) encryption \cite{8684919}.

Due to the vulnerabilities of IoT devices, the IoT verticals need to establish a secure MEC framework for cloud-assisted IoT environments and the secure APIs through which developers serve contents to such IoT applications of MEC \cite{8334918}. Extensive security surveys based on 5G properties in the areas of short-range IoT applications, delay-tolerant IoT applications, critical IoT application, and massive IoT applications are needed before 5G rollout \cite{8792139}. For example, 5G vertical security requirements of IoT-based electricity services within a smart city is needed before 5G roll out in UK \cite{8340149}. Attack vectors on SDN-based identity and access management, authentication, non-repudiation, audit, trust and assurance, compliance, confidentiality, integrity, availability, and privacy issues are to be considered, and proper safeguards and security risk mitigation techniques to support security at 5G access network, application layer, UE, management, core network, and infrastructure and virtualization components have to be deployed. The utilization of innovative security measures for IoT networks, such as two-factor authentication and key agreement schemes, can help in resisting various different types of attacks to ensure user privacy through both anonymity and unlinkability \cite{8267060}.
Since botnet is a major threat for IoT verticals \cite{8039305}, 5G roll-out design should be able to dynamically detect botnet traffic pattern and mitigate the attack in 5G network environment by leveraging the combination of SDN and NFV techniques to adapt botnet detection and reaction functions in 5G networks.

\subsection{Spectrum Coexistence and Trading}
The demands for new radio spectrum to meet the increasing growth of data traffic are rapidly increasing. To this end, the induction of new spectrum, location-based licensing of spectrum, efficient management of spectrum, and advent in green radio communication technologies are among the prominent solutions.


Dynamic spectrum sharing techniques can enable the sharing of the radio spectrum among two or more wireless systems and can effectively utilize the available radio frequencies \cite{SharmaFD2018}. The existing dynamic spectrum sharing models can be broadly categorized into the following three types: (i) Commons Model, (ii) Shared-use model, and (iii) exclusive-use model \cite{Hassan2017exclusive}.
In the spectrum commons model, radio spectrum is not owned by any provider and all the secondary users or unlicensed users can access the spectrum with equal rights. This sharing model suitable for spectrum sharing operation in the unlicensed bands such as Unlicensed National Information Infrastructure (U-NII),i.e., 5GHz, and Industrial, Scientific and Medical (ISM), i.e., 2.4 GHz. The main problem with this open access approach is that unlicensed users may suffer from the inter-user interference, and this may result in the network congestion.

In the shared-use model, the secondary users utilize the vacant spectrum or underutilized spectrum in an opportunistic way or an interference-avoidance manner without harming the normal operation of the primary users by using various Cognitive Radio (CR) techniques. The CR technology is considered as one promising technology to address the issue of spectrum scarcity, which can enable the coexistence of two or more wireless systems either in an opportunistic mode, i.e., interweave paradigm or with the interference avoidance mode, i.e., underlay paradigm \cite{Sharma2015CR}. The interweave paradigm mainly deals with the spectrum sharing or database assisted techniques, while the underlay techniques enable the spectrum sharing techniques of wireless systems by means of suitable interference mitigation techniques such as beamforming and power control.

Besides the aforementioned interweave and underlay paradigms, several advanced spectrum sharing mechanisms including Carrier Aggregation (CA) and Channel Bonding (CB) \cite{Khan2014CA}, Spectrum Access System (SAS), Licensed Shared Access (LSA) and  Licensed Assisted Access (LAA) \cite{Mukherjee2016LAA} have been investigated in the literature. Furthermore, various other spectrum sharing techniques such as spectrum leasing, spectrum trading, spectrum harvesting and spectrum mobility have been discussed in the literature towards improving the spectrum efficiency and energy efficiency of wireless networks \cite{Yang2016advanced}.

On the other hand, the secondary users can acquire the exclusive spectrum usages right for the required bandwidth and duration in the exclusive-use model. These exclusive rights can be obtained from the primary system either by purchasing the spectrum from the primary service providers or spectrum licensees or by providing a cooperation reward, for instance the relaying of the primary data. As compared to the shared-used model, this approach has several advantages for the secondary users as they do not need to sense the primary channel and do not need to switch from one channel to another channel. This mode of  spectrum sharing in the exclusive-mode is also called as spectrum trading \cite{Hassan2017exclusive}, which can be implemented either directly between the secondary and primary service providers or can be managed by a spectrum manager/broker or a spectrum exchange market.

An auction operation may be needed or not depending on the required duration for spectrum trading, i.e., long duration of spectrum trading needs an auction while a short duration may not need an auction. Auctioning process becomes more suitable in the scenarios with a high demand and limited supply since a seller can attract more benefits by involving multiple bidders. However, this method depends on various factors such as low number of bidders or a single bidder, all bidders asking below the required price, time dependency, and the need of multiple iterations to obtain a suitable solution, thus resulting to the need of a real-time multi-seller and multi-channel model while considering the dynamicity of the channel and traffic conditions \cite{Hassan2017exclusive}. On the other hand, non-auction based models can be designed based on game-theoretical models and may be categorized into monetary \cite{Bajaj2015spectrumtrading} and non-monetary types \cite{Simeone2008leasing}.

Considering the increasing demand for Ultra-Reliable and Low-Latency Communications (URLLC) applications such as industrial automation, there arises the need of meeting reliability requirements with the available spectrum. The existing factory automation generally utilizes unlicensed Industrial, Scientific and Medical (ISM) bands and benefits from their wider bandwidths in handling large traffic volumes. However, the reliability targets of below 10 ms can not meet with the existing solutions \cite{Holfeld2016factory}, and also the unlicensed mode of operation requires the careful design of minimizing or avoiding interference being subject to strict regulatory constraints. Considering these drawbacks of utilizing unlicensed bands or factory automation applications, the authors in \cite{8685776} demonstrated the possibility of utilizing 5G cellular licensed band as an alternative option for factory automation applications, and also showed the significance of utilizing integrated unlicensed and licensed bands in terms of economic viability. However, several challenges in terms of synchronized operation in the unlicensed and licensed bands and  over-the-air inter-system coordination needs to be addressed to realize this integrated spectrum utilization approach in practice.

The authors in \cite{8708839} discussed the use of 24 GHz as the Gigabit wireless networking spectrum based on the forecast methodology for 5G spectrum by K-ICT for IMT 2020 Korea \cite{KICTIMT2020Korea}, where three different forecast methodologies for 5G spectrum needs, namely, traffic forecast-based approach, technical performance-based approach, and application-based approach have been identified. From the analysis presented in \cite{8708839} regarding the use of 24 GHz, there arises the requirement of over 370 MHz spectrum for mobile broadband services for 1 Gbps speed and this requirement is expected to increase by 10\% till 2024. Also, the analysis pointed out the need of about 233,282 base stations in the analyzed regions consisting of dense urban, urban and sub-urban areas to support the increasing number of 5G users.
Moreover, understanding the spectrum usage pattern with the help of suitable spectrum monitoring techniques/platforms is a vital task towards enhancing the radio spectrum utilization efficiency of Beyond 5G dynamic sharing systems.  In this direction, Machine Learning (ML) techniques are of significant importance to predict the future spectrum usage and to address the inefficiency issues in spectrum management and utilization. The existing ML techniques can be broadly categorized into supervised, unsupervised and reinforcement learning \cite{Sharma2019mMTC}. Authors in \cite{7289481} carried out the analysis of spectrum occupancy in CR networks by utilizing different supervised and unsupervised ML techniques. Under the supervised learning approach, various techniques including Naive Bayesian Classifier (NBC), Support Vector Machine (SVM), Decision Trees (DT), and Linear Regression (LR) were considered while under the unsupervised learning, Hidden Markov Model (HMM) approach was investigated along with their numerical comparisons in terms of computational time and classification accuracy.

\begin{table}[t]
  \centering
  \caption{Summary of 5G networks target KPIs}
    \begin{tabular}{|p{2.8em}|p{3.1em}|p{2.5em}|p{2.7em}|p{2.7em}|p{2.4em}|p{2.98em}|}
    \hline
    \textbf{Service Platform} &
      \textbf{Device Density (/km$^2$)} &
      \multicolumn{1}{p{2.6em}|}{\textbf{Mob- ility (km/h)}} &
      \textbf{User Data Rate.} &
      \textbf{Cell Edge Rate} &
      \textbf{Lat-ency} &
      \textbf{ Availa-bility (Relia-bility)}
      \bigstrut\\
    \hline \hline
    \textbf{eMBB} &
      Up to 10k &
      \multicolumn{1}{l|}{\multirow{3}[6]{2.6em}{0 to 360}} &
      50 - 100 Mbps &
      50 Mbps DL, 25 Mbps UL &
      10 - 50ms &
      \multicolumn{1}{l|}{99\%}
      \bigstrut\\
\cline{1-2}\cline{4-7}    \textbf{mMTC} &
      Up to 1m &
       &
      Up to 100kbps &
      100 kbps DL and UL & $>$50ms &
      \multicolumn{1}{l|}{99\%}
      \bigstrut\\
\cline{1-2}\cline{4-7}
      \textbf{URLLC} & Up to 10k & \ & Up to 100 Mbps & 10 Mbps DL and UL & 1 - 50ms & 99.9\% - 99.999\%
      \bigstrut\\
    \hline
    \end{tabular}%
  \label{tab_5G_KPIs}%
\vspace{8pt}
  \centering
  \caption{Coverage range approximation for favourable channel conditions.}
    \begin{tabular}{|p{2em}|p{3.785em}|c|c|c|}
    \hline
    \multicolumn{1}{|p{2em}|}{\multirow{2}[4]{2em}{\textbf{Frequency Band}}} &
      \multirow{2}[4]{3.785em}{\textbf{Envir-onment}} &
      \multicolumn{3}{p{12.645em}|}{\textbf{Coverage range (kms)}}
      \bigstrut\\
\cline{3-5}     &
      \multicolumn{1}{c|}{} &
      \multicolumn{1}{p{4.215em}|}{\textbf{eMBB}} &
      \multicolumn{1}{p{4.215em}|}{\textbf{URLLC}} &
      \multicolumn{1}{p{4.215em}|}{\textbf{mMTC}}
      \bigstrut\\
    \hline
     &
      Rural &
      2.62 &
      2.69 &
      12.5
      \bigstrut\\
\cline{2-5}    \multicolumn{1}{|p{5.145em}|}{\textbf{700MHz}} &
      Sub-Urban &
      0.8 &
      0.82 &
      7
      \bigstrut\\
\cline{2-5}     &
      Urban &
      0.59 &
      0.65 &
      4.3
      \bigstrut\\
    \hline
     &
      Rural &
      0.62 &
      0.65 &
      5.65
      \bigstrut\\
\cline{2-5}    \multicolumn{1}{|p{5.145em}|}{\textbf{3.5GHz}} &
      Sub-Urban &
      0.17 &
      0.17 &
      2.09
      \bigstrut\\
\cline{2-5}     &
      Urban &
      0.09 &
      0.09 &
      0.47
      \bigstrut\\
    \hline
     &
      Rural &
      0.16 &
      0.17 &
      1.52
      \bigstrut\\
\cline{2-5}    \multicolumn{1}{|p{5.145em}|}{\textbf{26GHz}} &
      Sub-Urban &
      0.13 &
      0.13 &
      0.97
      \bigstrut\\
\cline{2-5}     &
      Urban &
      0.08 &
      0.08 &
      0.48
      \bigstrut\\
    \hline
    \end{tabular}%
  \label{table_converage_range_bands}%
\end{table}

Furthermore, a conceptual framework of end-to-end learning for spectrum monitoring applications has been presented in \cite{Kulin2018endtoend} along with a generic methodology to design and implement wireless signal classifiers followed by two case-studies related to modulation recognition and wireless technology interference detection. The end-to-end learning concept investigated in \cite{Kulin2018endtoend} refers to a learning procedure in which the features of a wireless signal are extracted, and a wireless signal classifier is utilized to classify the received signals. Moreover, it may not be reliable in practice to learn the radio spectrum usage by an individual node due to several issues such as multi-path fading and hidden node problem, leading to the need of collaborative learning and spectrum sharing strategy \cite{SharmaPIMRC2017}. Also, a non-collaborative way of spectrum usage learning in mmWave bands may result in challenging issues due to the involved directional antenna beams, and this can be improved by enabling collaboration among the secondary users to predict or estimate the spectrum occupancy distribution of the radio channels \cite{Liglobecomlearning}.


\section{case study: A Typical UK City}

This section presents a case study conducted to represent a typical UK city. For this purpose, the local authority involved typical West Midlands city layout is used. The study is focused on to demonstrate the potential of street furniture and public building for infrastructure sharing.
There are four major MNOs in the UK. In Table \ref{tab_UK_consumers}, the details of current major infrastructure consumers in the UK are provided, which are classified into MNOs, MVNOs, private networks, and semi-private networks classes.

We begin with the introduction of the 5G testbed environment and experimental setup, then further proceed with conducting a comprehensive analysis on our findings.

\begin{figure}[t]
  \centering
  \includegraphics[width=\columnwidth]{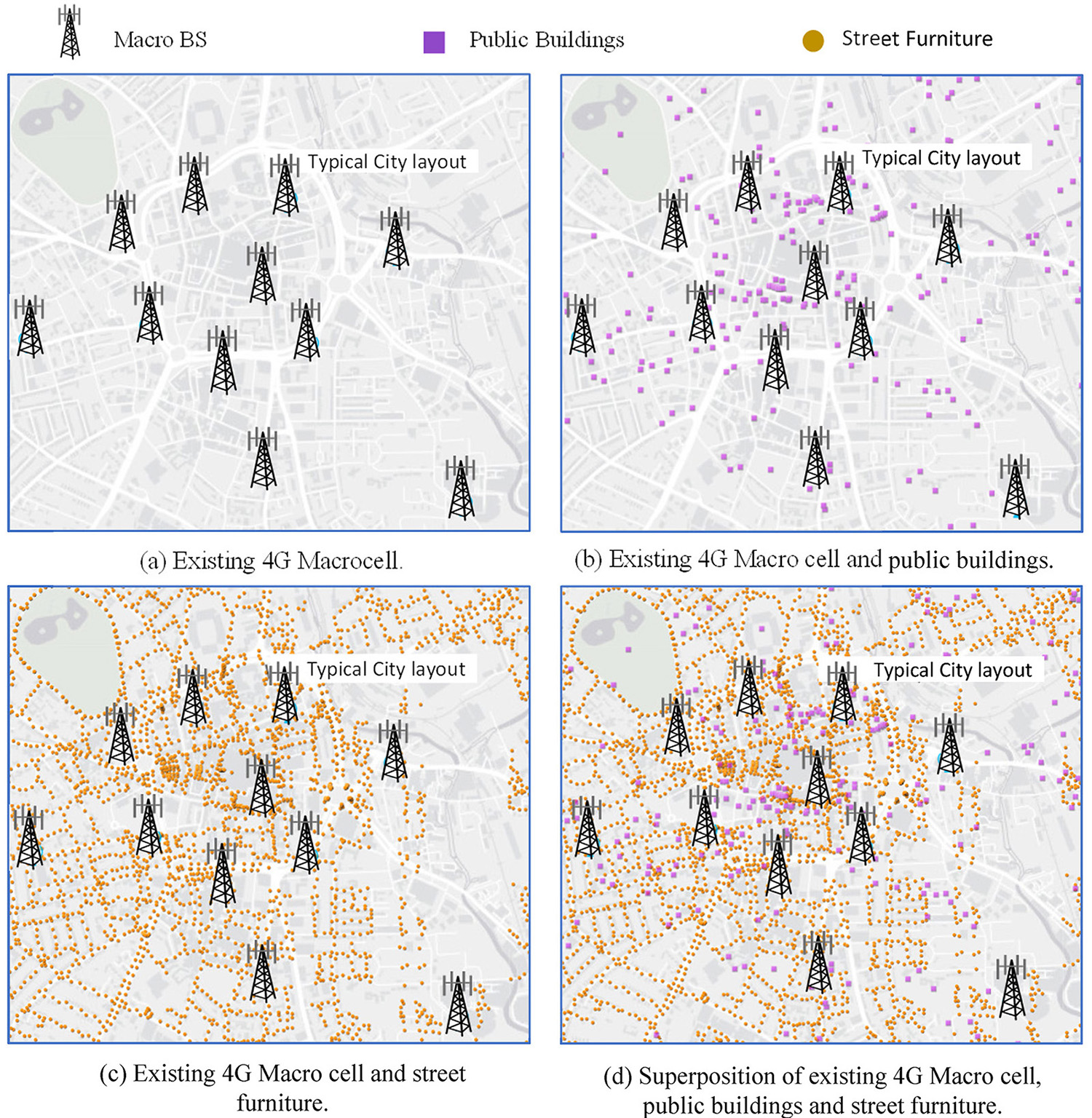}
  \caption{Case study-- A 5 km$^2$ area of a typical city.}
  \label{fig_case_s13}
\end{figure}

\begin{table}[ht]
  \centering
  \caption{Details of the infrastructure consumers.}
    \begin{tabular}{|p{4em}|p{8.5em}|p{13.5em}|}
    \hline
    \textbf{MNO } &
      \textbf{MVNO } &
      \textbf{Private \& semi private network}
      \bigstrut\\
    \hline\hline
    \multirow{2}[2]{*}{EE} &
      \multirow{2}[2]{9.4em}{Virgin Mobile; ASDA Mobile; BT Mobile } &
       National Roads Telecommunications Services (NRTS) and Traffic Scotland
      \bigstrut[t]\\
    \multicolumn{1}{|c|}{} &
      \multicolumn{1}{c|}{} &
      Network Rail Telecom
      \bigstrut[b]\\
\hline    Three  &
      iDmobile; FreedomPop; The People's Operator  &
      Airwave/ESN
      \bigstrut\\
\hline    O2  &
      Tesco Mobile; Lycamobile; Giffgaff  &
      Power Utilities
      \bigstrut\\
\hline   Vodafone &
      Lebara Mobile; TalkTalk Mobile; TalkMobile  &
       Sigfox/Arqiva
      \bigstrut\\
    \hline
    \end{tabular}%
  \label{tab_UK_consumers}
\end{table}%

\subsection{5G Testbed Environment and Experimental Setup}

West-Midland 5G (WM5G) is the UK's largest 5G testbed, which is available at the 5G business \& innovation center (5GBIC) in Birmingham City University (BCU) UK. The WM5G is a public-private partnership initiative with an investment of up to \textsterling 150m. This partnership aims at providing innovation to attain enhanced digital productivity and economy. The WM5G at 5GBIC is utilized for the conduction of the proposed case study.

The 5G testbed key performance indicators (KPIs) are summarized in Table \ref{tab_5G_KPIs}.
The derived coverage range estimate for different 5G target services (i.e., eMBB, URLLC, and mMTC) is provided in Table \ref{table_converage_range_bands}. The table exhibits a more ubiquitous view of the scale of infrastructural provisions required for 5G services. The characterization of coverage range into different cellular conditions/environments (i.e., rural, sub-urban, and urban) and various frequency bands (i.e., 700MHz, 3.5GHz, and 26GHz) is also presented.

The Case study is conducted for a 5 km$^2$ area of a typical UK city. The existing telecom and local authorities owned infrastructure in the considered region is plotted in Fig. \ref{fig_case_s13}. In Fig. \ref{fig_case_s13}(a), (b), (c), and (d), the location of existing 4G macro-cells, with added public buildings, with added street furniture, and superimposed all available infrastructure are plotted.
There is a requirement to provide good quality outdoor services for at least 140,000 premises to which the obligation holder does not currently provide the good coverage. Moreover, there is a requirement to deploy at least 500 new wide area mobile sites in rural areas, to be co-located at least at 1 km distance from existing sites. This concludes that each of the new 500 sites shall have a minimum EIRP of 43 dBm.  The conduction of a comprehensive study on ICNIRP requirements and environmental sustainability of the structures is suggested. It is observed that the potential of public buildings and street furniture to facilitate the necessary infrastructure for 5G deployment also provides a significant opportunity for the Local Authority to directly or indirectly participate in the business model along with the MNOs \cite{patwary2016universal}.

\begin{figure}[t]
  \centering
  \includegraphics[width=\columnwidth]{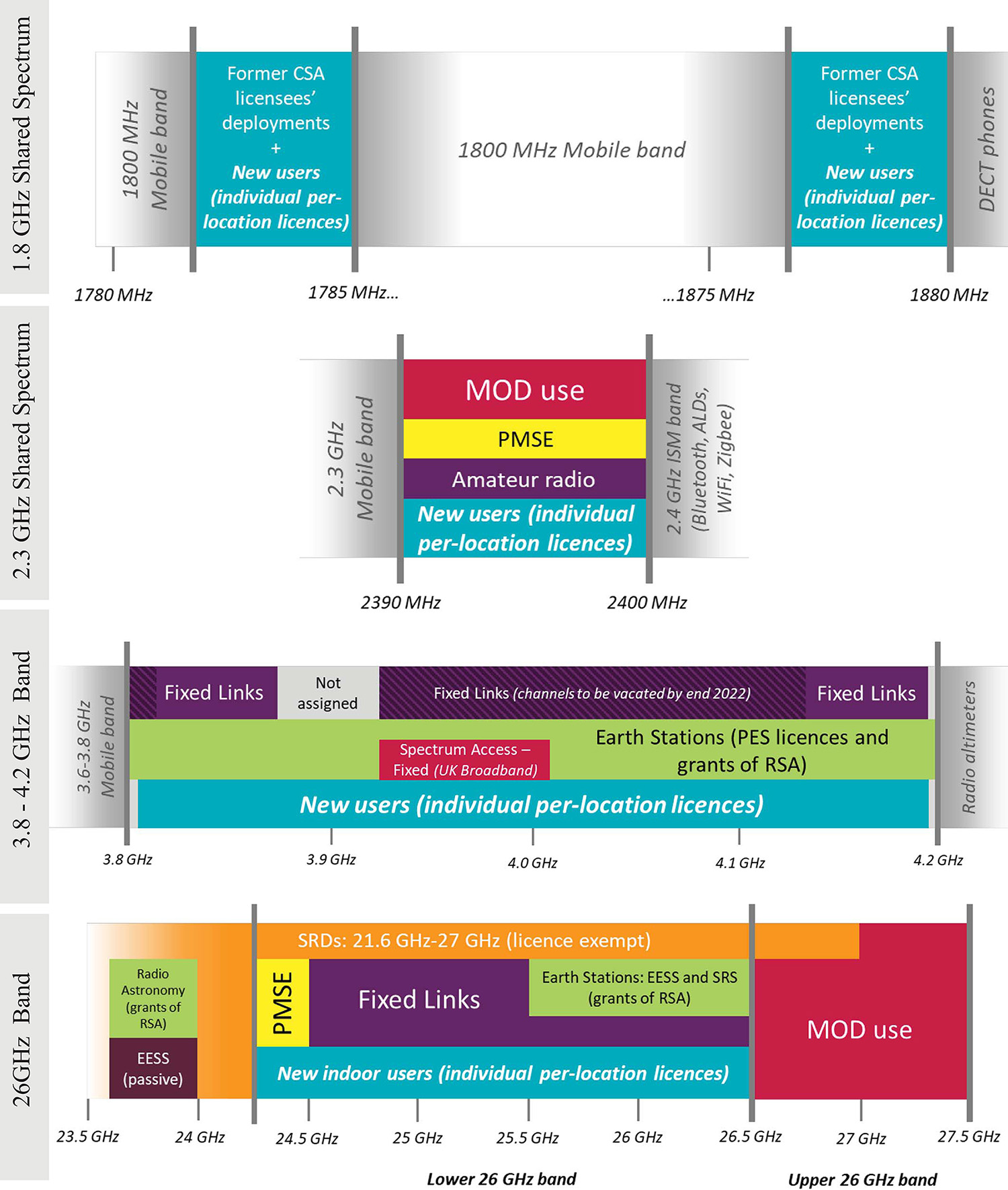}
  \caption{Ofcom 5G Spectrum sharing framework UK; i.e., for 1.8GHz, 2.3GHz, 3.8-4.2GHz, Lower-26GHz, and Upper-26GHz bands \cite{Ofcom_location_licensing_2019}.}
  \label{fig_spectrum_uk}
\end{figure}


\subsection{Spectrum Flexibility Policy in the UK}

The radio frequency spectrum being a core enabler of wireless communications has a high significance in shaping a country's economy and society. The huge deployment and radio spectrum costs are regarded as the potential delaying causes in the provision of 5G technology innovations.
Enabling opportunities for innovation with spectrum sharing has a strong potential in reducing the overall cost. In the light of growing interest in the use of communication applications and services introduced by 5G, there is a need to develop viable solutions for licensing the radio spectrum for meeting the local connectivity needs. In this regard, the location-based licensing of the radio spectrum can significantly help the MNOs in utilizing the 5G radio spectrum in the suitable local regions. The UK, in this regard, has become the first country in location-based spectrum licensing. The Ofcom has become the leader by releasing the location-based licensing of 5G compatible bands \cite{ofcom_licence}.
This new way of spectrum licensing also mainly opens the possibilities for location wise re-licensing of the radio spectrum which is allocated to the MNOs but it is not being utilized in the locations.
The nature of 5G-compatible radio, e.g., millimeter-wave (mmWave) signal propagation, and small-sized cells suit the adoption of such location-based licenses due to their shorter coverage distance from a base station.
The offering of location-based spectrum licensing will also open opportunities for small drivers (businesses, organizations, enterprises, industries, etc) to set up their own customized local wireless network at a cheaper cost with higher reliability and security provisions. The extended application scenarios of this arrangement may include private wireless networks for machine-to-machine communications in industrial, agricultural, others for various useful services. Moreover, the deployment of setup for wireless broadband connectivity in rural areas using fixed wireless access (FWA) may also benefit from it.

\begin{figure}
  \centering
  \includegraphics[width=\columnwidth]{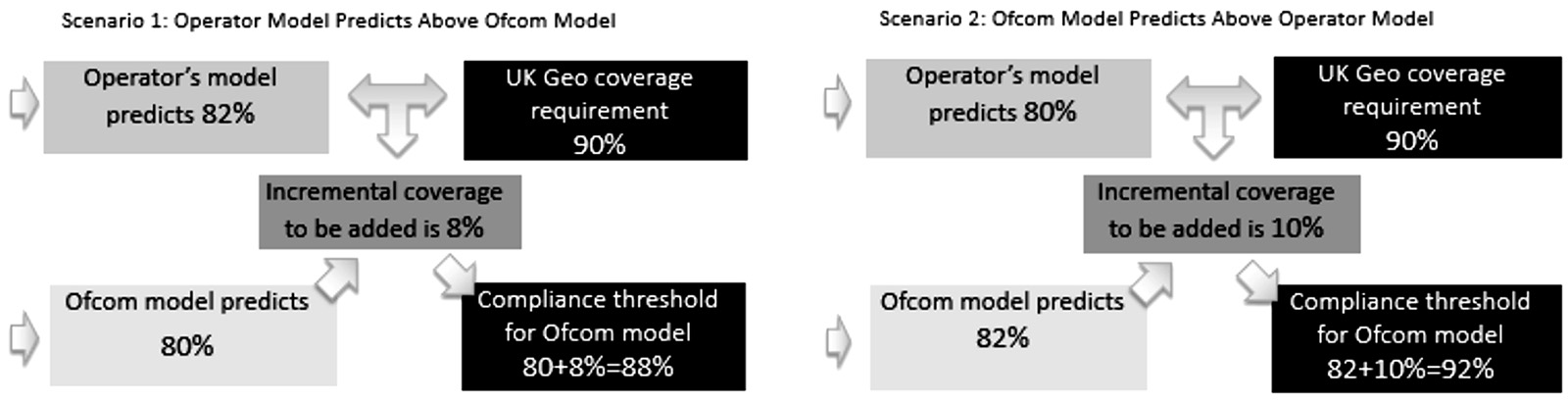}
  \caption{Scenarios for Ofcom and operator model prediction using UK's geo coverage model and requirements \cite{ofcom_licence,Ofcom_location_licensing_2019}.}
  \label{scenarios_ofcome_flow_UK}
\end{figure}

The framework of four prime 5G bands for location-based shared licensing released by Ofcom UK \cite{Ofcom_location_licensing_2019} are shown in Fig. \ref{fig_spectrum_uk}, i.e., 1.8GHz, 2.3GHz, 3.8 - 4.2 GHz, and 26GHz. The configuration of existing users in the corresponding bands are also indicated. The provision of a new regulatory framework for new users to access the spectrum is provided under \emph{Mobile Trading Regulations} \cite{Ofcom_location_licensing_2011_ammeded}.
The radio spectrum landscape indicating the new users with individual per-location licences, fixed links licences, concurrent spectrum access (CSA) licences, and former CSA licences are also indicated in the Fig. \ref{fig_spectrum_uk}.
The scenarios for Ofcom and operator model prediction using UK's geo coverage model and requirements are illustrated in Fig. \ref{scenarios_ofcome_flow_UK}. The compliance threshold for Ofcom model is 88\% and 92\% for the scenario of operator-model predicting above Ofcome-model and Ofcome-model predicting above operator-model, respectively.
The summary of prices of channels of different sizes in the UK by Ofcom \cite{Ofcom_location_licensing_2019} is provided in Table \ref{table_channel_price_UK}.
The average size of channel is considered as 40 MHz, where the cost for channel sizes higher and lower than 40 MHz are decided in proportionate to that. Complete details of tariffs by Ofcom for the year 2019/2020 can be found in \cite{Ofcom_Tarrif_2019_20}.

\begin{table}[t]
  \centering
  \caption{Channel size and price in the UK \cite{Ofcom_location_licensing_2019}}
    \begin{tabular}{|c|c|}
    \hline
    \multicolumn{1}{|p{10.43em}|}{\textbf{Channel size (MHz)}} &
      \multicolumn{1}{p{10.43em}|}{\textbf{Price per channel ( \textsterling)}}
      \bigstrut\\
    \hline\hline
      2$\times$3.3 &
      80
      \bigstrut\\
    \hline
    10 &
      80
      \bigstrut\\
    \hline
    20 &
      160
      \bigstrut\\
    \hline
    30 &
      240
      \bigstrut\\
    \hline
    40 &
      320
      \bigstrut\\
    \hline
    50 &
      400
      \bigstrut\\
    \hline
    60 &
      480
      \bigstrut\\
    \hline
    80 &
      640
      \bigstrut\\
    \hline
    100 &
      800
      \bigstrut\\
    \hline
    \end{tabular}%
  \label{table_channel_price_UK}%
\end{table}%

\begin{table}[t]
  \centering
  \caption{Typical infrastructure capability for different network architecture site types}
    \begin{tabular}{|p{5.215em}|p{3.115em}|p{3.315em}|p{3.115em}|p{3.85em}|p{2.7em}|}
    \hline
    \textbf{Site Deployment Option:} &
      \textbf{Macro Cell} &
      \textbf{Microcell} &
      \textbf{Picocell} &
      \textbf{Femtocell} &
      \textbf{Small Cell}
      \bigstrut\\
    \hline\hline
    \textbf{Greenfield} &
      \cmark &
      \xmark &
      \xmark &
      \xmark &
      \xmark
      \bigstrut\\
    \hline
    \textbf{Rooftop} &
      \cmark &
      \cmark &
      \xmark &
      \xmark &
      \xmark
      \bigstrut\\
    \hline
    \textbf{Streetworks} &
      \xmark &
      \cmark &
      \cmark &
      \xmark &
      \cmark
      \bigstrut\\
    \hline
    \textbf{Indoor} &
      \xmark &
      \xmark &
      \cmark &
      \cmark &
      \cmark
      \bigstrut\\
    \hline
    \end{tabular}%
  \label{table_cell_types}%
\end{table}%

\begin{table}[t]
  \centering
  \caption{Typical CapEx requirement to upgrade existing macro-cells with 5G capabilities. The sharing and non-sharing based costs comparison is presented. The cost heads shared between two MNOs are shaded in Gray color.}
    \begin{tabular}{|p{10em}|r|r|r|}
    \hline
    \multirow{2}[2]{*}{\textbf{Macro Item costs}} &
      \multicolumn{1}{c|}{\textbf{Not Shared}} &
      \multicolumn{2}{p{9em}|}{\textbf{Cost shared between two MNOs}}
      \bigstrut[b]\\
\cline{2-4}    \multicolumn{1}{|c|}{} &
      \multicolumn{1}{p{5em}|}{\textbf{Urban/Rural Sites  (Rooftop/ greenfield) (\textsterling)}} &
      \multicolumn{1}{p{4.3em}|}{\textbf{MNO 1 (\textsterling)}} &
      \multicolumn{1}{p{4.3em}|}{\textbf{MNO 2 (\textsterling)}}
      \bigstrut[t] \\
      \hline\hline
    \rowcolor[rgb]{ .851,  .851,  .851} Survey and Design &
      \cellcolor[rgb]{ 1,  1,  1}1,700.00 &
                 850.00  &
                 850.00
      \bigstrut[b]\\
    \hline
    \rowcolor[rgb]{ .851,  .851,  .851} Site Acquisition and planning &
      \cellcolor[rgb]{ 1,  1,  1}4,000.00 &
             2,000.00  &
              2,000.00
      \bigstrut\\
    \hline
    \rowcolor[rgb]{ .851,  .851,  .851} Civils works - Urban (mainly RT) &
      \cellcolor[rgb]{ 1,  1,  1}60,000.00 &
           30,000.00  &
            30,000.00
      \bigstrut\\
    \hline
    \rowcolor[rgb]{ .851,  .851,  .851} PSU &
      \cellcolor[rgb]{ 1,  1,  1}2,400.00 &
             1,200.00  &
              1,200.00
      \bigstrut\\
    \hline
    \rowcolor[rgb]{ .851,  .851,  .851} HVAC &
      \cellcolor[rgb]{ 1,  1,  1}6,600.00 &
             3,300.00  &
              3,300.00
      \bigstrut\\
    \hline
    Rigging &
      6,000.00 &
             6,000.00  &
              6,000.00
      \bigstrut\\
    \hline
    \rowcolor[rgb]{ .851,  .851,  .851} Antenna's (x6) &
      \cellcolor[rgb]{ 1,  1,  1}6,000.00 &
             3,000.00  &
              3,000.00
      \bigstrut\\
    \hline
    Antenna MIMO X3 (Based on \textsterling15k per 64x mMIMO) &
      45,000.00 &
           45,000.00  &
            45,000.00
      \bigstrut\\
    \hline
    Radio Hardware  &
      5,000 &
             5,000.00  &
              5,000.00
      \bigstrut\\
    \hline
    DICI &
      4,000.00 &
             4,000.00  &
              4,000.00
      \bigstrut\\
    \hline
    Transfer to Operations &
      1,000.00 &
             1,000.00  &
              1,000.00
      \bigstrut\\
    \hline
    Project Management &
      4,000.00 &
             4,000.00  &
              4,000.00
      \bigstrut\\
    \hline
    \textbf{Total without mMIMO} &
      \textbf{100,700.00} &
      \textbf{60,350.00 } &
      \textbf{60,350.00 }
      \bigstrut\\
    \hline
    \textbf{Total With mMIMO} &
      \textbf{145,700.00} &
      \textbf{105,350.00 } &
      \textbf{105,350.00 }
      \bigstrut\\
    \hline
    \textbf{Total without mMimo - including Risk and Margin} &
      \textbf{120,840.00} &
      \textbf{72,420.00 } &
      \textbf{72,420.00 }
      \bigstrut\\
    \hline
    \end{tabular}%
  \label{table_macrocell_items}%
\end{table}%

\subsection{Market Interests and Cost Analysis}

The overall 5G rollout cost in the UK is estimated as \textsterling 30bn -  \textsterling 50bn, while the UK mobile operator annual CapEx is estimated as \textsterling 2.5bn. Such a high cost of the rollout in the UK is highly unlikely to be solely supported by the MNOs.
This section discusses the economic constraints related with to rollout of 5G and B5G network in the UK. Moreover, the consumer market saturated and flat revenue prospects are also discussed.

There will be 518,345 sites required to be deployed in the UK \cite{NIC_report_2016}. These sites are classified as:(i) 7,616 sites for dense urban areas, (ii) 186,732 sites for urban areas, (iii) 309,014 sites for sub-urban areas, and (iv) 15,000 sites for rural areas.
Typical infrastructure capability for different site deployment options and cell types are indicated in Table \ref{table_cell_types}. The following subsections thoroughly discuss the CapEx and OpEx associated with new standalone and shared-infrastructure based macro-cell and small-cells.
\paragraph{Macro-cells}
Out of the total number of required sites, about 40,000 existing cell sites can be reused for macro-cells. The CapEx requirements to upgrade the existing macro-cells with 5G capability, without sharing the costs, for urban and rural areas are indicated in Table \ref{table_macrocell_items}.
Moreover, for the case of shared costs between different MNOs, the CapEx required for upgrading the existing macro-cells with 5G capabilities are also indicated in Table \ref{table_macrocell_items}. From different heads of CapEx, a few can be identified as shareable among multiple MNOs. In this context, the equal cost sharing map between two different MNOs (i.e., MNO1 and MNO2) is shown in the table with the rows representing the shared-cost heads shaded in Grey color. These items for macro-cells include, survey and design, site acquisition and planning, civil works, PSU, HVAC, and antennas. Moreover, the cost of non-shareable heads is also indicated.
The sharing of costs between two MNOs can reduce the cost from \textsterling 100,700.00 to \textsterling 60,350.00 and \textsterling 145,700.00 to \textsterling 105,350.00 for the cases of without and with mMIMO capability, respectively.  Moreover, by including the risk and margin costs, the cost of without mMIMO capability scenario can be reduced from \textsterling 120,840.00 to \textsterling 72,420.00 through sharing the potential head costs equally between two MNOs.

\begin{table*}[t]
  \centering
  \caption{Generic CapEx requirements for different deployment options for small-cells}
    \begin{tabular}{|c|c|c|c|c|p{5em}|}
    \hline
    \multicolumn{1}{|c|}{\multirow{2}[3]{*}{\textbf{Small-Cell items}}} &
      \multicolumn{2}{c|}{\textbf{Single Operator}} &
      \multicolumn{2}{c|}{\textbf{Shared between two operators}} &
      \multirow{2}[3]{*}{Sharing (\%)}
      \bigstrut
      \\
\cline{2-5}
    &
      \multicolumn{1}{p{7em}|}{\textbf{Deploying new pole. Cost (\textsterling)}} &
      \multicolumn{1}{p{7.5em}|}{\textbf{Using existing street furniture. Cost (\textsterling)}} &
      \multicolumn{1}{p{7.5em}|}{\textbf{Deploying new pole. Cost for each operator (\textsterling)}} &
      \multicolumn{1}{p{9.5em}|}{\textbf{Using existing street furniture. Cost for each operator (\textsterling)}} &
      \multicolumn{1}{c|}{}
      \bigstrut[t]\\
      \hline\hline
    \multicolumn{1}{|p{15em}|}{Design development of street-side pole (new)} &
      \multicolumn{1}{c|}{5000} &
      0 &
      2500 &
      0 &
      \multicolumn{1}{c|}{50\%}
      \bigstrut[b]\\
    \hline
    \multicolumn{1}{|p{13.645em}|}{Design \& Engineering (existing)} &
      \multicolumn{1}{c|}{2000} &
      2000 &
      1200 &
      1200 &
      \multicolumn{1}{c|}{60\%}
      \bigstrut\\
    \hline
    \multicolumn{1}{|p{13.645em}|}{Civil Work} &
      \multicolumn{1}{c|}{1000} &
      1000 &
      800 &
      800 &
      \multicolumn{1}{c|}{80\%}
      \bigstrut\\
    \hline
    \multicolumn{1}{|p{13.645em}|}{Power} &
      \multicolumn{1}{c|}{1000} &
      1000 &
      500 &
      500 &
      \multicolumn{1}{c|}{50\%}
      \bigstrut\\
    \hline
    \multicolumn{1}{|p{13.645em}|}{New Fibre (could be reduced if fibre already present)} &
      \multicolumn{1}{c|}{2660} &
      2660 &
      1330 &
      1330 &
      \multicolumn{1}{c|}{50\%}
      \bigstrut\\
    \hline
    \multicolumn{1}{|l|}{\multirow{2}[4]{*}{RF equipment ( non-shared)}} &
      \multicolumn{1}{c|}{\multirow{2}[4]{*}{7500}} &
      \multirow{2}[4]{*}{7500} &
      7500 &
      7500 &
      \multicolumn{1}{r|}{}
      \bigstrut\\
\cline{4-6}    \multicolumn{1}{|l|}{} &
      \multicolumn{1}{c|}{} &
       &
      \multicolumn{1}{p{7.5em}|}{(4500, if dual band transceiver is used of \textsterling 9000)} &
      \multicolumn{1}{p{7.5em}|}{(4500, if dual band transceiver is used of \textsterling9000)} &
      \multicolumn{1}{r|}{}
      \bigstrut\\
    \hline
    \multicolumn{1}{|p{13.645em}|}{\textbf{Total:}} &
      \multicolumn{1}{c|}{\textbf{19,160}} &
      \textbf{14,160} &
      \textbf{13,830} &
      \textbf{11,330} &
      Single band  antenna
      \bigstrut[t]\\ \hline
    \multicolumn{3}{|c|}{\multirow{2}[2]{*}{}} &
      \multicolumn{1}{p{7.5em}|}{Exiting pole + antenna sharing - 9000)} &
      \multicolumn{1}{p{7.5em}|}{Exiting pole + antenna sharing - 9000)} &
      \multicolumn{1}{r|}{}
      \bigstrut[b]\\
\cline{4-6}    \multicolumn{3}{|c|}{} &
      \textbf{10,830} &
      \textbf{8,330} &
      With dual band antenna
      \bigstrut[t]\\ \hline
    \end{tabular}%
  \label{table_Capex_smallcell}%
\end{table*}%

\paragraph{Small-cells}

The CapEx requirements for small-cells deployment with and without sharing options are presented in Table \ref{table_Capex_smallcell}. For the case of the sole operator, the costs of deploying a small-cell with a new pole compared to the cost for utilizing the existing street-furniture are presented. Moreover, for the case of sharing the small-cell poles between two operators, the cost comparison for new pole and street-furniture based deployment cases is also presented. For all four cases, the costs of design development of street-side pole, design and engineering aspects, civil work, power, new fibers, and RF-equipment items are discussed. Moreover, the anticipated sharing percentage of different items are also indicated.

For the case of sole operator BSs (no active sharing), the CapEx requirements for 5G small-cell deployment is calculated to be \textsterling 19,160 and \textsterling 14,160 for the cases of new-pole and street-furniture utilization based implementations, respectively.
For the model of cost-sharing between two operators, the CapEx requirements for 5G small-cell deployment is calculated to be \textsterling 13,830 and \textsterling 11,330 for the case of new-pole and street-furniture utilization based deployments, respectively. The costs are for single-band antenna BSs, whereas, the dual-band antenna based BSs deployment costs are also indicated.

To achieve the cost reduction offered by these models, the assessment of characteristics of available street furniture and public infrastructure in the UK for their capacity to hold extra weight, requisite height compliance, wind sustainability, neighborhood infrastructure availability, and suitability in terms of OpEx is essential. In Table \ref{table_small_cell_charac}, the target range for small-cell deployment compared to the available range of the public infrastructure in the UK in terms of discussed critical parameters are presented.

\begin{table}[t]
  \centering
  \caption{Characteristics of public infrastructure in the UK related to small cell deployment}
    \begin{tabular}{|p{10em}|p{5em}|p{9.0em}|}
    \hline
    \textbf{Parameter to consider} &
      \textbf{Target } &
      \textbf{Available range}
      \bigstrut\\
    \hline\hline
    Height &
      4 meter &
       6 - 10m
      \bigstrut\\
    \hline
    Weight bearing capacity  &
      7.5kg &
      5 - 15kg
      \bigstrut\\
    \hline
    Wind sustainability  &
      20km/h &
      22km/h for continuous 10 minutes
      \bigstrut\\
    \hline
    Neighbourhood (infrastructure distance) &
       50 - 80m &
       40 - 100m
      \bigstrut\\
    \hline
    Opex  &
       unknown &
       \textsterling 700 - 800 per pole per year
      \bigstrut\\
    \hline
    \end{tabular}%
  \label{table_small_cell_charac}
\end{table}%





\subsection{Revenue and Data Flow Model}

A shift in the attitude of the mobile service providers from transaction to relationship, marketing push to consumer pull, customer acquisition to customer retention, average revenue per user to average profit per user, intelligence in platform to intelligence in user equipment, investment infrastructure to leveraging key assets, and technology to content/data is presumed to arise.

The drive for revenue generation from 5G technologies can be achieved by devising separate short- and long-term strategies. In the short-term, the existing practices can be potentially evolved to offer the necessary infrastructure for attractive 5G business models; e.g., Neutral Hosting for small-cell sites, etc. To this end, the active and passive infrastructure sharing may be vital to facilitate the initial rollout. There is also a need to thoroughly study the available avenues for further reducing both capital expenditures (CapEx) and operational expenditures (OpEx). However, with the advent of 5G features, the existing backbone revenue-generating services (e.g., voice and text messaging) may not stay as an attractive proposition for long. In the long-term, big data analytics and innovative new services based entirely new platforms for value extraction may be strongly be required to build prevailing business models.

The notable items for consideration under the CapEx head can be listed as RF design and planning, site engineering, cabinet/antenna, baseband radios, installation and swap, project management, software (SW) license, cell-site gateway, antennas, site acquisition, power, backhaul, network implementation, and system integration. The primary items under OpEx head can be arranged as site rentals, power supply, backhaul, annual fees (SW etc.), network optimizations, central operations, hardware (HW) maintenance, SW Maintenance, and support setup. To attain the optimized CapEx and OpEx for the 5G network rollout, the following multi-step way forward is suggested,

\begin{itemize}
  \item Split the intended coverage area into small cells -- although the appropriate coverage can be attained with macro cell infrastructure.
  \item Use street furniture as possible infrastructure for small cells as the first option, followed by public and private buildings or rooftops and then telegraph poles if available.
  \item If possible, share antennas of the neighbouring spectrum at the same small cell site.
  \item Share fibre, power and other maintenance CapEx and OpEx.
\end{itemize}

\begin{figure}[t]
  \centering
  \includegraphics[width=0.84\columnwidth]{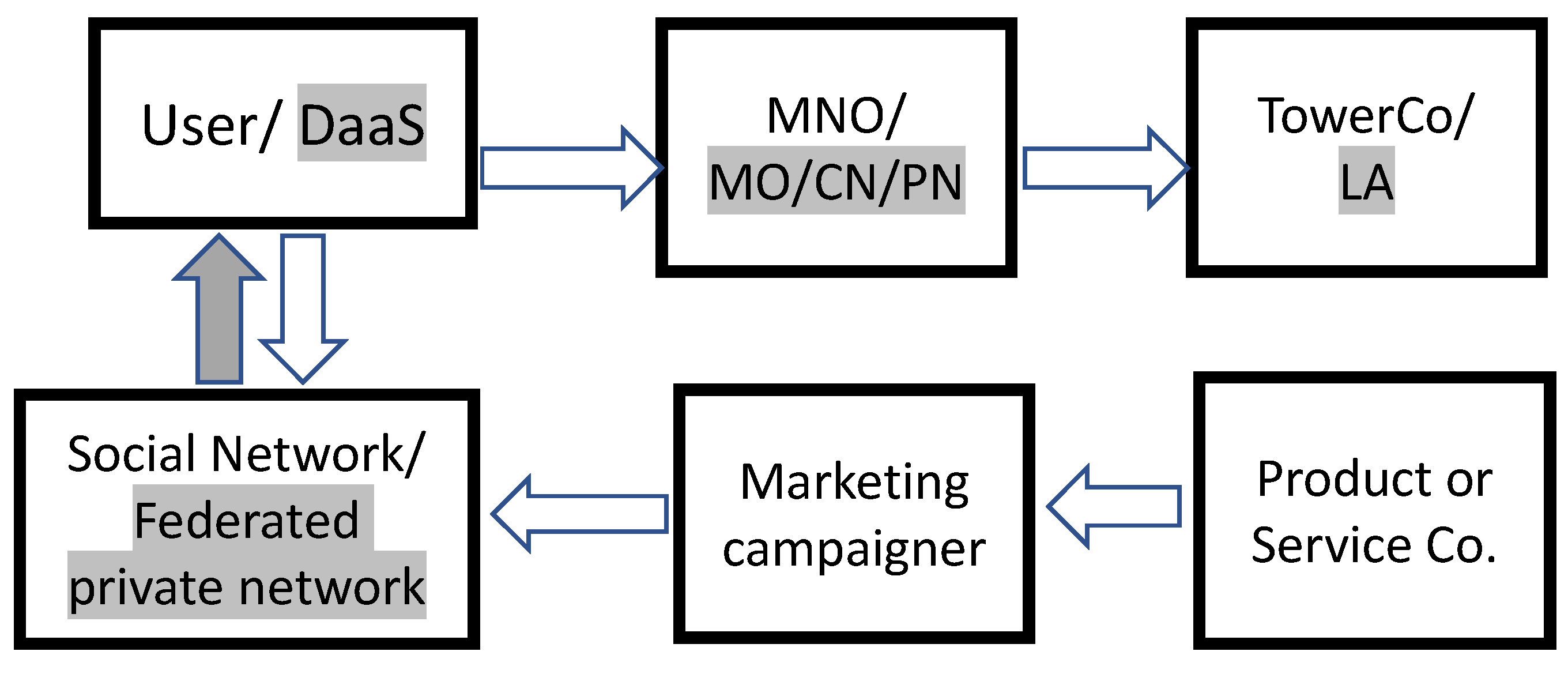}
  \caption{A simplified new revenue flow landscape with consumer, connectivity and service provider 5G will promote}\label{fig_s4_revenue_flow}
\end{figure}

There exist various strong synergies between infrastructure designs, business models, and revenue generation methods. Fig. \ref{fig_s4_revenue_flow} shows the primary landscape of revenue flow; where the consumer, network connectivity, and service providers act as the central elements of the ecosystem. The new market entrant are highlighted in gray color. The notable stakeholders and revenue-flow aspect are indicated in Table \ref{table_business_stakeholders}. The landscape of business is inscribed as infrastructure provider, connectivity provider, connectivity dependent services provider, and service consumers with mobility and connectivity. Examples of essential stakeholders, along with the crucial new entrants, are also quoted. There exist a substantial potential for the local authorities to become stakeholders in the business model by offering the public infrastructure as utilizable in the telecommunication setup deployments. The business model may adapt direct revenue sharing or utility-based incentives for inducing the local authorities into the future telecom ecosystem. The understanding of the potential benefits that the infrastructure sharing agreements can bring to the local authorities is also of critical importance for achieving long-term sustainability.

\begin{figure}[t]
  \centering
  \includegraphics[width=\columnwidth]{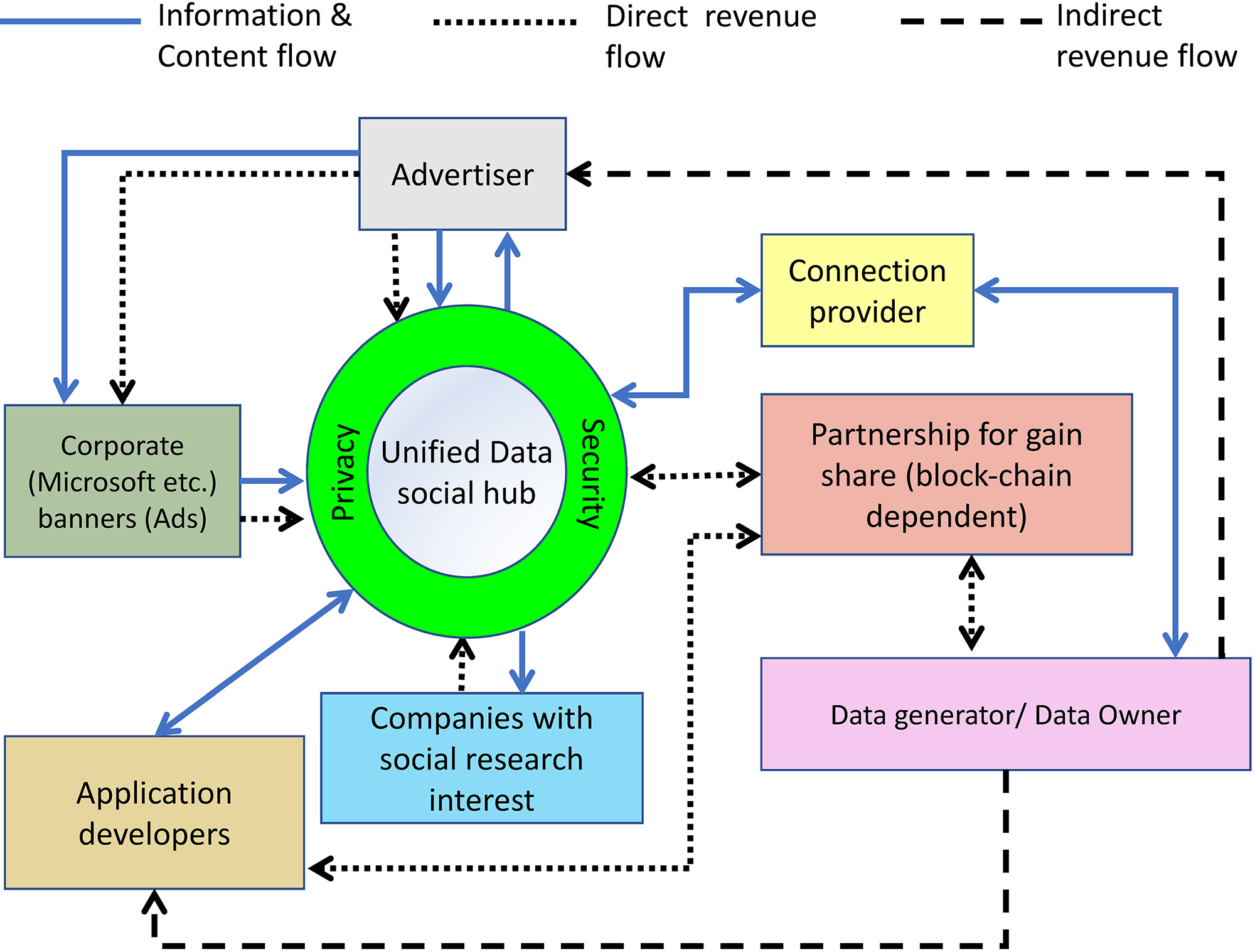}
  \caption{Proposed System Model for Data and Revenue Flow.}
  \label{fig_schematic_model}
\end{figure}

The proposed system model for data and revenue flow between different stakeholders is illustrated in Fig. \ref{fig_schematic_model}. A \emph{unified data special hub} may have two-way information and context flow with connection providers, advertisers, and companies with social research interests. The data-sharing may be protected with privacy and security suits. A controlled interface of applications developers to the data hub is critical in facilitating the development of advanced applications and indirect revenue flow, while also protecting the sanctity of the data. The application developers also require a two-way direct and in-direct revenue flow model with the partnership for gain share and data owners/generators, respectively. The partnership for gain share may be implemented as block-chain dependant facilitated through the direct revenue flow with application developers. The modeling and designing of the information-sharing platform between connection providers and data owners is another necessity to sanction the essential availability of data to the \emph{unified data social hub}. Information flow from advertisers to the data hub can run through corporate (Microsoft etc.) banners/advertisements while modeling a direct revenue flow mechanism between the two is also crucial. The companies with social research interest along with information sharing also need a revenue flow model to benefit from and to the unified data social hub.

The proposed data economy-oriented business model indicates the potential commodification of data and data transactions along with low-cost physical infrastructure and spectrum. It can be foreseen that the 5G network will introduce significant disruption within the Telco business ecosystem. Although there are large investment saving, we also understand, and considerations need to be made that not owning the physical infrastructure by telecom service providers there are potential legal complexities to acquire the positions and installations of the equipment on these proposed locations. This is due to heterogeneous public or private ownership of these infrastructures, and it is challenging to make a standard legal framework and financial model to acquire these resources. However, if we make a cost-benefit analysis, this is still a viable route for success without making this 5G infrastructure building a huge question for public-private investors.

\begin{table}[t]
  \centering
  \caption{The new-look; stakeholders and revenue-flow.}
    \begin{tabular}{|p{6.5em}|p{9em}|p{9em}|}
    \hline
    \textbf{Nature of business} &
      \textbf{ Stakeholder} &
      \cellcolor[rgb]{ .949,  .949,  .949}\textbf{New Entrant }
      \bigstrut\\
    \hline
    \textbf{Infrastructure provider} &
      TowerCo (4 in the UK, one of them dominates the market) &
      \cellcolor[rgb]{ .949,  .949,  .949}Local Authority (LA) as neutral host with public infrastructure offer to MNO
      \bigstrut\\
    \hline
    \textbf{Connectivity provider} &
      MNOs (4 in the UK) &
      \cellcolor[rgb]{ .949,  .949,  .949} Private network, community networks, micro operator with cheap infrastructure and spectrum
      \bigstrut\\
    \hline
    \textbf{Connectivity dependent service provider} &
      Social networks (Facebook, Google, Uber etc) &
      \cellcolor[rgb]{ .949,  .949,  .949} Federated private networks/  SME's providing data as a services decision as a service.
      \bigstrut\\
    \hline
    \textbf{Service consumer with Mobility and Connectivity} &
      User &
      \cellcolor[rgb]{ .949,  .949,  .949}Independent user/ SME's consuming data as a services decision as a service.
      \bigstrut\\
    \hline
    \end{tabular}%
  \label{table_business_stakeholders}
\end{table}%

\subsection{Future Recommendations}

In the following, we provide some recommendations towards accelerating the 5G deployment process and reducing its cost.

\begin{enumerate}

\item The local authority-owned street furniture and other associated public infrastructure assets form the strongest possible set of candidate infrastructure assets for 5G deployment. This provides Local Authorities with potential opportunities for creating new direct or indirect streams of revenue generation.

\item Considering the currently available radio technology and existing roadside infrastructure, it is possible to continue with the current models of active and passive infrastructure sharing.

\item Neutral Hosting may potentially disrupt the current models of TowerCo business oligopoly.

\item Data as a service and decision as a service is to be one of the prime revenue generating services to corporate and retail consumers for the 5G's success.

\item MNOs are required to adopt a harmonised co-existence with micro-operators, community network, and other private networks.

\item The success of federated private networks will introduce the potential of distributed web, a way forward to redefine the internet.

\end{enumerate}

\section{Conclusion}\label{Sec_4}

A thorough analysis of the potential long- and short-term transformative impacts anticipated from the 5G rollout has been conducted in this paper. The huge anticipated cost of 5G deployment is one of the major barriers in fully receiving the benefits from the innovative 5G communication technologies. Moreover, the lack of confidence of the MNOs on the revenue generation opportunities and existing business models is a primary determinant restraining them from investing the requisite deployment cost. To this end, the sharing of network infrastructure, public infrastructure, radio spectrum, and data are recommended as potential measures to reduce the deployment, operational, and maintenance costs as well as to develop a marketable 5G business model. The local authorities can potentially avail this opportunity to become direct or indirect partners in the telecommunication business model by offering the provisions of sharing the public infrastructure (street furniture, public buildings, etc.) for 5G deployment. A data sharing based value generation model as a long-term 5G business solution has also been proposed in this manuscript. The barriers in the sharing of data have been highlighted.

Moreover, the concerns associated with data privacy and security, along with their potential solutions, have also been studied. Based on the proposed resolutions, a 5G testbed environment-based case study for a typical UK city has been conducted. It has been concluded that public infrastructure sharing can potentially contribute to a reduction of 40-60\% in the deployment cost compared to the anticipated cost. In addition, the location-based shared spectrum licensing and proposed data value generation based long-term sustainable business model have been shown to help in further reducing the CapEx and OpEx significantly. Based on the conducted case study and analysis, a list of recommendations is proposed to reduce the 5G deployment cost and encourage the business.

\bibliographystyle{IEEEtran}
\bibliography{References}

\begin{thebibliography}{100}
\providecommand{\url}[1]{#1}
\csname url@samestyle\endcsname
\providecommand{\newblock}{\relax}
\providecommand{\bibinfo}[2]{#2}
\providecommand{\BIBentrySTDinterwordspacing}{\spaceskip=0pt\relax}
\providecommand{\BIBentryALTinterwordstretchfactor}{4}
\providecommand{\BIBentryALTinterwordspacing}{\spaceskip=\fontdimen2\font plus
\BIBentryALTinterwordstretchfactor\fontdimen3\font minus
  \fontdimen4\font\relax}
\providecommand{\BIBforeignlanguage}[2]{{%
\expandafter\ifx\csname l@#1\endcsname\relax
\typeout{** WARNING: IEEEtran.bst: No hyphenation pattern has been}%
\typeout{** loaded for the language `#1'. Using the pattern for}%
\typeout{** the default language instead.}%
\else
\language=\csname l@#1\endcsname
\fi
#2}}
\providecommand{\BIBdecl}{\relax}
\BIBdecl

\bibitem{3GPPrel15}
{Technical Specification Group Services and Systems Aspects}, ``{System
  Architecture for the {5G} System; Stage 2, Releases 15},'' {3GPP}, Tech. Rep.
  {document 3GPP TS 23.501, V15.2.0}, 2018.

\bibitem{obiodu20175g}
E.~Obiodu and M.~Giles, ``The {5G} era: age of boundless connectivity and
  intelligent automation,'' \emph{GSM Association}, 2017.

\bibitem{7414384}
M.~{Agiwal}, A.~{Roy}, and N.~{Saxena}, ``Next generation {5G} wireless
  networks: A comprehensive survey,'' \emph{IEEE Communications Surveys
  Tutorials}, vol.~18, no.~3, pp. 1617--1655, thirdquarter 2016.

\bibitem{shafi20175g}
M.~Shafi, A.~F. Molisch, P.~J. Smith, T.~Haustein, P.~Zhu, P.~D. Silva,
  F.~Tufvesson, A.~Benjebbour, and G.~Wunder, ``{5G: A tutorial overview of
  standards, trials, challenges, deployment, and practice},'' \emph{IEEE J.
  Sel. Areas Commun.}, vol.~35, no.~6, pp. 1201--1221, Jun. 2017.

\bibitem{Simsek2016tactile}
M.~{Simsek}, A.~{Aijaz}, M.~{Dohler}, J.~{Sachs}, and G.~{Fettweis},
  ``{5G}-enabled tactile internet,'' \emph{IEEE J. Sel. Areas Commun.},
  vol.~34, no.~3, pp. 460--473, Mar. 2016.

\bibitem{schneir2019business}
J.~R. Schneir, A.~Ajibulu, K.~Konstantinou, J.~Bradford, G.~Zimmermann,
  H.~Droste, and R.~Canto, ``A business case for {5G} mobile broadband in a
  dense urban area,'' \emph{Telecommunications Policy}, Aug. 2019.

\bibitem{wisely2018capacity}
D.~Wisely, N.~Wang, and R.~Tafazolli, ``Capacity and costs for {5G} networks in
  dense urban areas,'' \emph{IET Communications}, vol.~12, no.~19, pp.
  2502--2510, 2018.

\bibitem{8412482}
K.~{David} and H.~{Berndt}, ``{6G} vision and requirements: Is there any need
  for beyond {5G?}'' \emph{IEEE Veh. Technol. Mag.}, vol.~13, no.~3, pp.
  72--80, Sep. 2018.

\bibitem{chianiopen}
M.~Chiani, E.~Paolini, and F.~Callegati, ``Open issues and beyond {5G},''
  \emph{{5G} Italy White eBook: from Research to Market}, Nov. 2018.

\bibitem{QML_6G_Junaid}
S.~J. {Nawaz}, S.~K. {Sharma}, S.~{Wyne}, M.~N. {Patwary}, and
  M.~{Asaduzzaman}, ``Quantum machine learning for {6G} communication networks:
  {State}-of-the-art and vision for the future,'' \emph{IEEE Access}, vol.~7,
  pp. 46\,317--46\,350, Apr. 2019.

\bibitem{Saad_6G_2019}
W.~Saad, M.~Bennis, and M.~Chen, ``A vision of {6G} wireless systems:
  Applications, trends, technologies, and open research problems,'' \emph{arXiv
  preprint arXiv:1902.10265}, Mar. 2019.

\bibitem{Tariq6G_2019}
F.~Tariq, M.~R.~A. Khandaker, K.-K. Wong, M.~Imran, M.~Bennis, , and M.~Debbah,
  ``A speculative study on {6G},'' \emph{arXiv preprint arXiv:1902.06700v1},
  Feb. 2019.

\bibitem{8760275}
Z.~{Baiqing}, C.~{Fan}, X.~{Wang}, X.~{Duan}, B.~{Wang}, and J.~{Wang}, ``6g
  technologies: Key drivers, core requirements, system architectures, and
  enabling technologies,'' \emph{IEEE Vehicular Technology Magazine}, vol.~PP,
  pp. 1--1, Jul 2019.

\bibitem{8782879}
P.~{Yang}, Y.~{Xiao}, M.~{Xiao}, and S.~{Li}, ``6g wireless communications:
  Vision and potential techniques,'' \emph{IEEE Network}, vol.~33, no.~4, pp.
  70--75, Jul. 2019.

\bibitem{Report_MobileEconomy2019}
``The mobile economy,'' GSMA, Report, 2019.

\bibitem{8300495}
F.~{Mekuria} and L.~{Mfupe}, ``Spectrum sharing \& affordable broadband in
  {5G},'' in \emph{proc. of Global Wireless Summit (GWS)}, Oct. 2017, pp.
  114--118.

\bibitem{8516977}
W.~{Xie}, N.~{Mao}, and K.~{Rundberget}, ``Cost comparisons of backhaul
  transport technologies for {5G} fixed wireless access,'' in \emph{proc. of
  IEEE 5G World Forum (5GWF)}, Jul. 2018, pp. 159--163.

\bibitem{oughton2019assessing}
E.~J. Oughton, Z.~Frias, S.~V.~D. Gaast, and R.~V.~D. Berg, ``Assessing the
  capacity, coverage and cost of {5G} infrastructure strategies: Analysis of
  the netherlands,'' \emph{Telematics and Informatics}, vol.~37, pp. 50--69,
  2019.

\bibitem{jones2019commentary}
P.~Jones and D.~Comfort, ``A commentary on the rollout of {5G} mobile in the
  uk,'' \emph{Journal of Public Affairs}, p. e1993, 2019.

\bibitem{yaghoubi2019techno}
F.~Yaghoubi, M.~Mahloo, L.~Wosinska, P.~Monti, F.~S. Farias, J.~C. W.~A. Costa,
  and J.~Chen, \emph{Techno-economic and Business Feasibility Analysis of {5G}
  Transport Networks}.\hskip 1em plus 0.5em minus 0.4em\relax John Wiley \&
  Sons, Ltd, 2019, ch.~13, pp. 273--295.

\bibitem{oughton2018cost}
E.~J. Oughton and Z.~Frias, ``The cost, coverage and rollout implications of
  {5G} infrastructure in britain,'' \emph{Telecommunications Policy}, vol.~42,
  no.~8, pp. 636--652, 2018.

\bibitem{ofcom_licence}
\BIBentryALTinterwordspacing
Ofcom, ``Enabling wireless innovation through local licensing,'' July 2019.
  [Online]. Available:
  \url{https://www.ofcom.org.uk/consultations-and-statements/category-1/enabling-opportunities-for-innovation?showall=1}
\BIBentrySTDinterwordspacing

\bibitem{AndrewsJSAC5G}
J.~G. {Andrews}, S.~{Buzzi}, W.~{Choi}, S.~V. {Hanly}, A.~{Lozano}, A.~C.~K.
  {Soong}, and J.~C. {Zhang}, ``What will {5G} be?'' \emph{IEEE J. Sel. Areas
  Commun.}, vol.~32, no.~6, pp. 1065--1082, Jun. 2014.

\bibitem{8452975}
C.~{Li}, C.~{Li}, K.~{Hosseini}, S.~B. {Lee}, J.~{Jiang}, W.~{Chen}, G.~{Horn},
  T.~{Ji}, J.~E. {Smee}, and J.~{Li}, ``{5G}-based systems design for tactile
  internet,'' \emph{Proceedings of the IEEE}, vol. 107, no.~2, pp. 307--324,
  Feb. 2019.

\bibitem{8467353}
S.~F. {Abedin}, M.~G.~R. {Alam}, S.~M.~A. {Kazmi}, N.~H. {Tran}, D.~{Niyato},
  and C.~S. {Hong}, ``Resource allocation for ultra-reliable and enhanced
  mobile broadband iot applications in fog network,'' \emph{IEEE Transactions
  on Communications}, vol.~67, no.~1, pp. 489--502, Jan. 2019.

\bibitem{IMTvision}
{International Telecommunication Union}, ``{IMT Vision—Framework and Overall
  Objectives of the Future Development of IMT for 2020 and Beyond},'' IMT,
  Tech. Rep. {document M Series 2083}, Sep. 2015.

\bibitem{8403963}
H.~{Ji}, S.~{Park}, J.~{Yeo}, Y.~{Kim}, J.~{Lee}, and B.~{Shim},
  ``Ultra-reliable and low-latency communications in {5G} downlink: Physical
  layer aspects,'' \emph{IEEE Wireless Communications}, vol.~25, no.~3, pp.
  124--130, Jun. 2018.

\bibitem{Bockelmann2018mmtc}
C.~{Bockelmann}, N.~K. {Pratas}, G.~{Wunder}, S.~{Saur}, M.~{Navarro},
  D.~{Gregoratti}, G.~{Vivier}, E.~{De Carvalho}, and {\it et al.}, ``Towards
  massive connectivity support for scalable {mMTC} communications in {5G}
  networks,'' \emph{{IEEE Access}}, vol.~6, pp. 28\,969--28\,992, May 2018.

\bibitem{Sharmaedgecloudedge}
S.~K. {Sharma} and X.~{Wang}, ``Live data analytics with collaborative edge and
  cloud processing in wireless {IoT} networks,'' \emph{IEEE Access}, vol.~5,
  pp. 4621--4635, Mar. 2017.

\bibitem{6736746}
F.~{Boccardi}, R.~W. {Heath}, A.~{Lozano}, T.~L. {Marzetta}, and P.~{Popovski},
  ``Five disruptive technology directions for {5G},'' \emph{IEEE Communications
  Magazine}, vol.~52, no.~2, pp. 74--80, Feb. 2014.

\bibitem{sanfilippo2018concise}
G.~Sanfilippo, O.~Galinina, S.~Andreev, S.~Pizzi, and G.~Araniti, ``A concise
  review of {5G} new radio capabilities for directional access at {mmWave}
  frequencies,'' in \emph{Internet of Things, Smart Spaces, and Next Generation
  Networks and Systems}.\hskip 1em plus 0.5em minus 0.4em\relax Springer, 2018,
  pp. 340--354.

\bibitem{8594703}
F.~{Jameel}, S.~{Wyne}, S.~J. {Nawaz}, and Z.~{Chang}, ``Propagation channels
  for {mmWave} vehicular communications: State-of-the-art and future research
  directions,'' \emph{IEEE Wireless Communications}, vol.~26, no.~1, pp.
  144--150, Feb. 2019.

\bibitem{6736761}
E.~G. {Larsson}, O.~{Edfors}, F.~{Tufvesson}, and T.~L. {Marzetta}, ``Massive
  {MIMO} for next generation wireless systems,'' \emph{IEEE Communications
  Magazine}, vol.~52, no.~2, pp. 186--195, Feb. 2014.

\bibitem{5462882}
S.~J. {Nawaz}, B.~H. {Qureshi}, and N.~M. {Khan}, ``A generalized {3-D}
  scattering model for a macrocell environment with a directional antenna at
  the {BS},'' \emph{IEEE Transactions on Vehicular Technology}, vol.~59, no.~7,
  pp. 3193--3204, Sep. 2010.

\bibitem{mansoor2017massive}
B.~Mansoor, S.~J. Nawaz, and S.~Gulfam, ``{Massive-MIMO} sparse uplink channel
  estimation using implicit training and compressed sensing,'' \emph{Applied
  Sciences}, vol.~7, no.~1, p.~63, 2017.

\bibitem{dahlman20185g}
E.~Dahlman, S.~Parkvall, and J.~Skold, \emph{5G NR: The next generation
  wireless access technology}.\hskip 1em plus 0.5em minus 0.4em\relax Academic
  Press, 2018.

\bibitem{8357810}
L.~{Dai}, B.~{Wang}, Z.~{Ding}, Z.~{Wang}, S.~{Chen}, and L.~{Hanzo}, ``A
  survey of non-orthogonal multiple access for {5G},'' \emph{IEEE
  Communications Surveys Tutorials}, vol.~20, no.~3, pp. 2294--2323,
  thirdquarter 2018.

\bibitem{7263349}
L.~{Dai}, B.~{Wang}, Y.~{Yuan}, S.~{Han}, C.~{I}, and Z.~{Wang},
  ``Non-orthogonal multiple access for {5G}: solutions, challenges,
  opportunities, and future research trends,'' \emph{IEEE Communications
  Magazine}, vol.~53, no.~9, pp. 74--81, Sep. 2015.

\bibitem{8396262}
X.~{Xia}, K.~{Xu}, Y.~{Wang}, and Y.~{Xu}, ``A {5G}-enabling technology:
  Benefits, feasibility, and limitations of in-band full-duplex {mMIMO},''
  \emph{IEEE Vehicular Technology Magazine}, vol.~13, no.~3, pp. 81--90, Sep.
  2018.

\bibitem{8516948}
J.~F. {Valenzuela-Valdes}, A.~{Palomares}, J.~C. {Gonzalez-Macias},
  A.~{Valenzuela-Valdes}, P.~{Padilla}, and F.~{Luna-Valero}, ``On the
  ultra-dense small cell deployment for {5G} networks,'' in \emph{2018 IEEE 5G
  World Forum (5GWF)}, Jul. 2018, pp. 369--372.

\bibitem{7422408}
X.~{Ge}, S.~{Tu}, G.~{Mao}, C.~{Wang}, and T.~{Han}, ``{5G} ultra-dense
  cellular networks,'' \emph{IEEE Wireless Communications}, vol.~23, no.~1, pp.
  72--79, Feb. 2016.

\bibitem{7994617}
A.~{Basta}, A.~{Blenk}, K.~{Hoffmann}, H.~J. {Morper}, M.~{Hoffmann}, and
  W.~{Kellerer}, ``Towards a cost optimal design for a {5G} mobile core network
  based on {SDN} and {NFV},'' \emph{IEEE Transactions on Network and Service
  Management}, vol.~14, no.~4, pp. 1061--1075, Dec. 2017.

\bibitem{8320765}
I.~{Afolabi}, T.~{Taleb}, K.~{Samdanis}, A.~{Ksentini}, and H.~{Flinck},
  ``Network slicing and softwarization: A survey on principles, enabling
  technologies, and solutions,'' \emph{IEEE Communications Surveys Tutorials},
  vol.~20, no.~3, pp. 2429--2453, thirdquarter 2018.

\bibitem{Abbas2018mec}
N.~{Abbas}, Y.~{Zhang}, A.~{Taherkordi}, and T.~{Skeie}, ``Mobile edge
  computing: A survey,'' \emph{IEEE Internet Things J.}, vol.~5, no.~1, pp.
  450--465, Feb. 2018.

\bibitem{7446253}
S.~{Buzzi}, C.~{I}, T.~E. {Klein}, H.~V. {Poor}, C.~{Yang}, and A.~{Zappone},
  ``A survey of energy-efficient techniques for {5G} networks and challenges
  ahead,'' \emph{IEEE J. Sel. Areas Commun.}, vol.~34, no.~4, pp. 697--709,
  Apr. 2016.

\bibitem{8304385}
N.~{Zhang}, P.~{Yang}, J.~{Ren}, D.~{Chen}, L.~{Yu}, and X.~{Shen}, ``Synergy
  of big data and {5G} wireless networks: Opportunities, approaches, and
  challenges,'' \emph{IEEE Wireless Communications}, vol.~25, no.~1, pp.
  12--18, Feb. 2018.

\bibitem{7429688}
Y.~{He}, F.~R. {Yu}, N.~{Zhao}, H.~{Yin}, H.~{Yao}, and R.~C. {Qiu}, ``Big data
  analytics in mobile cellular networks,'' \emph{IEEE Access}, vol.~4, pp.
  1985--1996, Mar. 2016.

\bibitem{7515114}
D.~E. {O'Leary}, ``Ethics for big data and analytics,'' \emph{IEEE Intelligent
  Systems}, vol.~31, no.~4, pp. 81--84, Jul. 2016.

\bibitem{8428412}
X.~{Jing}, Z.~{Yan}, and W.~{Pedrycz}, ``Security data collection and data
  analytics in the internet: A survey,'' \emph{IEEE Communications Surveys
  Tutorials}, vol.~21, no.~1, pp. 586--618, Firstquarter 2019.

\bibitem{ICNIRP_guideline_distance_98}
\BIBentryALTinterwordspacing
``{ICNIRP} guidelines for limiting exposure to time-varying electric, magnetic
  and electromagnetic fields,'' INTERNATIONAL COMMISSION ON NON-IONIZING
  RADIATION PROTECTION (ICNIRP), Technical Report, 1998. [Online]. Available:
  \url{https://www.icnirp.org/en/publications/index.html}
\BIBentrySTDinterwordspacing

\bibitem{SCF_19}
\BIBentryALTinterwordspacing
``{Small Cell Forum Ltd},'' Registered in the UK no. 6295097, Tech. Rep.,
  2007-2019. [Online]. Available: \url{https://www.smallcellforum.org/}
\BIBentrySTDinterwordspacing

\bibitem{iws_sol_1}
\BIBentryALTinterwordspacing
iWireless Solutions~(iWS), ``Small cell mast.'' [Online]. Available:
  \url{https://www.iwireless-solutions.com/}
\BIBentrySTDinterwordspacing

\bibitem{union2015imt}
I.~T.~U. ({ITU}), ``{IMT} traffic estimates for the years 2020 to 2030,''
  \emph{Report ITU-R M. 2370--0, ITU-R Radiocommunication Sector of ITU}, 2015.

\bibitem{8648405}
K.~{Valtanen}, J.~{Backman}, and S.~{Yrjola}, ``Blockchain-powered value
  creation in the {5G} and smart grid use cases,'' \emph{IEEE Access}, vol.~7,
  pp. 25\,690--25\,707, Feb. 2019.

\bibitem{7345407}
P.~{Schneider} and G.~{Horn}, ``Towards {5G} security,'' in \emph{2015 IEEE
  Trustcom/BigDataSE/ISPA}, vol.~1, Aug. 2015, pp. 1165--1170.

\bibitem{8334918}
I.~{Ahmad}, T.~{Kumar}, M.~{Liyanage}, J.~{Okwuibe}, M.~{Ylianttila}, and
  A.~{Gurtov}, ``Overview of {5G} security challenges and solutions,''
  \emph{IEEE Communications Standards Magazine}, vol.~2, no.~1, pp. 36--43,
  Mar. 2018.

\bibitem{7081072}
X.~{Duan} and X.~{Wang}, ``Authentication handover and privacy protection in
  {5G} hetnets using software-defined networking,'' \emph{IEEE Communications
  Magazine}, vol.~53, no.~4, pp. 28--35, Apr. 2015.

\bibitem{8125684}
D.~{Fang}, Y.~{Qian}, and R.~Q. {Hu}, ``Security for {5G} mobile wireless
  networks,'' \emph{IEEE Access}, vol.~6, pp. 4850--4874, 2018.

\bibitem{8706883}
A.~{Braeken}, M.~{Liyanage}, P.~{Kumar}, and J.~{Murphy}, ``Novel {5G}
  authentication protocol to improve the resistance against active attacks and
  malicious serving networks,'' \emph{IEEE Access}, vol.~7, pp.
  64\,040--64\,052, 2019.

\bibitem{7842433}
Z.~{Ding}, Y.~{Liu}, J.~{Choi}, Q.~{Sun}, M.~{Elkashlan}, C.~{I}, and H.~V.
  {Poor}, ``Application of non-orthogonal multiple access in lte and {5G}
  networks,'' \emph{IEEE Communications Magazine}, vol.~55, no.~2, pp.
  185--191, Feb. 2017.

\bibitem{8792139}
R.~{Khan}, P.~{Kumar}, D.~N.~K. {Jayakody}, and M.~{Liyanage}, ``A survey on
  security and privacy of {5G} technologies: Potential solutions, recent
  advancements and future directions,'' \emph{IEEE Communications Surveys
  Tutorials}, pp. 1--1, Aug. 2019.

\bibitem{8684919}
M.~{Karmoose}, C.~{Fragouli}, S.~{Diggavi}, R.~{Misoczki}, L.~L. {Yang}, and
  Z.~{Zhang}, ``Using {mm-Waves} for secret key establishment,'' \emph{IEEE
  Communications Letters}, vol.~23, no.~6, pp. 1077--1080, Jun. 2019.

\bibitem{8523804}
I.~{Abdulqadder}, D.~{Zou}, I.~{Aziz}, B.~{Yuan}, and W.~{Dai}, ``Deployment of
  robust security scheme in {SDN} based {5G} network over {NFV} enabled cloud
  environment,'' \emph{IEEE Transactions on Emerging Topics in Computing}, pp.
  1--1, Nov. 2018.

\bibitem{8335294}
J.~{Wang}, Y.~{Huang}, S.~{Jin}, R.~{Schober}, X.~{You}, and C.~{Zhao},
  ``Resource management for device-to-device communication: A physical layer
  security perspective,'' \emph{IEEE Journal on Selected Areas in
  Communications}, vol.~36, no.~4, pp. 946--960, Apr. 2018.

\bibitem{7954591}
F.~{Tian}, P.~{Zhang}, and Z.~{Yan}, ``A survey on {C-RAN} security,''
  \emph{IEEE Access}, vol.~5, pp. 13\,372--13\,386, Jun. 2017.

\bibitem{8403769}
M.~{Lichtman}, R.~{Rao}, V.~{Marojevic}, J.~{Reed}, and R.~P. {Jover}, ``{5G
  NR} jamming, spoofing, and sniffing: Threat assessment and mitigation,'' in
  \emph{2018 IEEE International Conference on Communications Workshops (ICC
  Workshops)}, May 2018, pp. 1--6.

\bibitem{8314666}
J.~{Ni}, X.~{Lin}, and X.~S. {Shen}, ``Efficient and secure service-oriented
  authentication supporting network slicing for {5G}-enabled {IoT},''
  \emph{IEEE Journal on Selected Areas in Communications}, vol.~36, no.~3, pp.
  644--657, Mar. 2018.

\bibitem{8260929}
J.~{Backman}, S.~{Yrjola}, K.~{Valtanen}, and O.~{Mammela}, ``Blockchain
  network slice broker in {5G: Slice} leasing in factory of the future use
  case,'' in \emph{proc. of Internet of Things Business Models, Users, and
  Networks}, Nov. 2017, pp. 1--8.

\bibitem{8712553}
I.~{Ahmad}, S.~{Shahabuddin}, T.~{Kumar}, J.~{Okwuibe}, A.~{Gurtov}, and
  M.~{Ylianttila}, ``Security for {5G} and beyond,'' \emph{IEEE Communications
  Surveys Tutorials}, pp. 1--1, May 2019.

\bibitem{8771320}
H.~C. {Rudolph}, A.~{Kunz}, L.~L. {Iacono}, and H.~V. {Nguyen}, ``Security
  challenges of the {3GPP 5G} service based architecture,'' \emph{IEEE
  Communications Standards Magazine}, vol.~3, no.~1, pp. 60--65, Mar. 2019.

\bibitem{8335293}
X.~{Lu}, D.~{Niyato}, N.~{Privault}, H.~{Jiang}, and P.~{Wang}, ``Managing
  physical layer security in wireless cellular networks: A cyber insurance
  approach,'' \emph{IEEE Journal on Selected Areas in Communications}, vol.~36,
  no.~7, pp. 1648--1661, Jul. 2018.

\bibitem{jover2019current}
R.~P. Jover, ``The current state of affairs in {5G} security and the main
  remaining security challenges,'' \emph{arXiv preprint arXiv:1904.08394},
  2019.

\bibitem{8454666}
K.~{Xiao}, W.~{Li}, M.~{Kadoch}, and C.~{Li}, ``On the secrecy capacity of {5G
  MmWave} small cell networks,'' \emph{IEEE Wireless Communications}, vol.~25,
  no.~4, pp. 47--51, Aug. 2018.

\bibitem{8685867}
V.~{Messie}, G.~{Fromentoux}, X.~{Marjou}, and N.~L. {Omnes}, ``{BALAdIN} for
  blockchain-based {5G} networks,'' in \emph{proc. of Conference on Innovation
  in Clouds, Internet and Networks and Workshops (ICIN)}, Feb. 2019, pp.
  201--205.

\bibitem{8470085}
Z.~{Chen}, S.~{Chen}, H.~{Xu}, and B.~{Hu}, ``A security authentication scheme
  of {5G} ultra-dense network based on block chain,'' \emph{IEEE Access},
  vol.~6, pp. 55\,372--55\,379, Sep. 2018.

\bibitem{8731639}
H.~{Dai}, Z.~{Zheng}, and Y.~{Zhang}, ``Blockchain for internet of things: A
  survey,'' \emph{IEEE Internet of Things Journal}, pp. 1--1, 2019.

\bibitem{8320551}
K.~{Fan}, Y.~{Ren}, Y.~{Wang}, H.~{Li}, and Y.~{Yang}, ``Blockchain-based
  efficient privacy preserving and data sharing scheme of content-centric
  network in {5G},'' \emph{IET Communications}, vol.~12, no.~5, pp. 527--532,
  Mar. 2018.

\bibitem{8356363}
S.~{Kiyomoto}, A.~{Basu}, M.~S. {Rahman}, and S.~{Ruj}, ``On blockchain-based
  authorization architecture for {beyond-5G} mobile services,'' in \emph{2017
  12th International Conference for Internet Technology and Secured
  Transactions (ICITST)}, Dec. 2017, pp. 136--141.

\bibitem{8707070}
B.~{Nour}, A.~{Ksentini}, N.~{Herbaut}, P.~A. {Frangoudis}, and H.~{Moungla},
  ``A blockchain-based network slice broker for {5G} services,'' \emph{IEEE
  Networking Letters}, vol.~1, no.~3, pp. 99--102, Sep. 2019.

\bibitem{8368983}
K.~{Valtanen}, J.~{Backman}, and S.~{Yrjola}, ``Creating value through
  blockchain powered resource configurations: Analysis of {5G} network slice
  brokering case,'' in \emph{2018 IEEE Wireless Communications and Networking
  Conference Workshops (WCNCW)}, Apr. 2018, pp. 185--190.

\bibitem{8726067}
Y.~{Dai}, D.~{Xu}, S.~{Maharjan}, Z.~{Chen}, Q.~{He}, and Y.~{Zhang},
  ``Blockchain and deep reinforcement learning empowered intelligent {5G}
  beyond,'' \emph{IEEE Network}, vol.~33, no.~3, pp. 10--17, May 2019.

\bibitem{8304873}
J.~{Li}, Z.~{Zhao}, and R.~{Li}, ``Machine learning-based ids for
  software-defined {5G} network,'' \emph{IET Networks}, vol.~7, no.~2, pp.
  53--60, Mar. 2018.

\bibitem{8283694}
L.~{Fernandez Maimo}, A.~L. {Perales Gomez}, F.~J. {Garcia Clemente}, M.~{Gil
  Perez}, and G.~{Martinez Perez}, ``A self-adaptive deep learning-based system
  for anomaly detection in {5G} networks,'' \emph{IEEE Access}, vol.~6, pp.
  7700--7712, 2018.

\bibitem{8781929}
J.~{Cao}, P.~{Yu}, X.~{Xiang}, M.~{Ma}, and H.~{Li}, ``Anti-quantum fast
  authentication and data transmission scheme for massive devices in {5G
  NB-IoT} system,'' \emph{IEEE Internet of Things Journal}, pp. 1--1, Jul.
  2019.

\bibitem{8276259}
K.~{Lee}, S.~{Lee}, C.~{Seo}, and K.~{Yim}, ``{TRNG (True Random Number
  Generator)} method using visible spectrum for secure communication on {5G}
  network,'' \emph{IEEE Access}, vol.~6, pp. 12\,838--12\,847, Jan. 2018.

\bibitem{8315003}
B.~{Bordel}, A.~B. {Orúe}, R.~{Alcarria}, and D.~{Sánchez-De-Rivera}, ``An
  intra-slice security solution for emerging {5G} networks based on
  pseudo-random number generators,'' \emph{IEEE Access}, vol.~6, pp.
  16\,149--16\,164, Mar. 2018.

\bibitem{8718355}
S.~{Garg}, K.~{Kaur}, G.~{Kaddoum}, S.~H. {Ahmed}, and D.~N.~K. {Jayakody},
  ``{SDN} based secure and privacy-preserving scheme for vehicular networks: A
  {5G} perspective,'' \emph{IEEE Transactions on Vehicular Technology}, pp.
  1--1, May 2019.

\bibitem{8268052}
J.~{Liu}, L.~{Zhang}, R.~{Sun}, X.~{Du}, and M.~{Guizani}, ``Mutual
  heterogeneous signcryption schemes for {5G} network slicings,'' \emph{IEEE
  Access}, vol.~6, pp. 7854--7863, 2018.

\bibitem{8088619}
X.~{Zhang}, A.~{Kunz}, and S.~{Schröder}, ``Overview of {5G} security in
  {3GPP},'' in \emph{proc. of IEEE Conference on Standards for Communications
  and Networking}, Sep. 2017, pp. 181--186.

\bibitem{8343866}
V.~{Ortega}, F.~{Bouchmal}, and J.~F. {Monserrat}, ``Trusted {5G} vehicular
  networks: Blockchains and content-centric networking,'' \emph{IEEE Vehicular
  Technology Magazine}, vol.~13, no.~2, pp. 121--127, Jun. 2018.

\bibitem{7433471}
M.~{Hashem Eiza}, Q.~{Ni}, and Q.~{Shi}, ``Secure and privacy-aware
  cloud-assisted video reporting service in 5g-enabled vehicular networks,''
  \emph{IEEE Transactions on Vehicular Technology}, vol.~65, no.~10, pp.
  7868--7881, Oct. 2016.

\bibitem{8701642}
L.~{Xie}, Y.~{Ding}, H.~{Yang}, and X.~{Wang}, ``Blockchain-based secure and
  trustworthy internet of things in {SDN}-enabled {5G-VANETs},'' \emph{IEEE
  Access}, vol.~7, pp. 56\,656--56\,666, Apr. 2019.

\bibitem{7939143}
A.~{Hussein}, I.~H. {Elhajj}, A.~{Chehab}, and A.~{Kayssi}, ``{SDN VANETs in
  5G:} an architecture for resilient security services,'' in \emph{proc. of
  International Conference on Software Defined Systems (SDS)}, May 2017, pp.
  67--74.

\bibitem{8340149}
G.~{Arfaoui}, P.~{Bisson}, R.~{Blom}, R.~{Borgaonkar}, H.~{Englund},
  E.~{Felix}, F.~{Klaedtke}, P.~K. {Nakarmi}, M.~{Naslund}, P.~{O'H{an}lon},
  J.~{Papay}, J.~{Suomalainen}, M.~{Surridge}, J.~{Wary}, and A.~{Zahariev},
  ``A security architecture for {5G} networks,'' \emph{IEEE Access}, vol.~6,
  pp. 22\,466--22\,479, 2018.

\bibitem{8267060}
S.~{Shin} and T.~{Kwon}, ``Two-factor authenticated key agreement supporting
  unlinkability in 5g-integrated wireless sensor networks,'' \emph{IEEE
  Access}, vol.~6, pp. 11\,229--11\,241, Jan. 2018.

\bibitem{8039305}
M.~G. {Perez}, A.~H. {Celdran}, F.~{Ippoliti}, P.~G. {Giardina}, G.~{Bernini},
  R.~M. {Alaez}, E.~{Chirivella-Perez}, F.~J.~G. {Clemente}, G.~M. {Perez},
  E.~{Kraja}, G.~{Carrozzo}, J.~M.~A. {Calero}, and Q.~{Wang}, ``Dynamic
  reconfiguration in {5G} mobile networks to proactively detect and mitigate
  botnets,'' \emph{IEEE Internet Computing}, vol.~21, no.~5, pp. 28--36, Sep.
  2017.

\bibitem{SharmaFD2018}
S.~K. {Sharma}, T.~E. {Bogale}, L.~B. {Le}, S.~{Chatzinotas}, X.~{Wang}, and
  B.~{Ottersten}, ``Dynamic spectrum sharing in {5G} wireless networks with
  full-duplex technology: Recent advances and research challenges,'' \emph{IEEE
  Commun. Surveys Tuts.}, vol.~20, no.~1, pp. 674--707, Feb. 2018.

\bibitem{Hassan2017exclusive}
M.~R. {Hassan}, G.~C. {Karmakar}, J.~{Kamruzzaman}, and B.~{Srinivasan},
  ``Exclusive use spectrum access trading models in cognitive radio networks: A
  survey,'' \emph{IEEE Communications Surveys Tutorials}, vol.~19, no.~4, pp.
  2192--2231, Fourthquarter 2017.

\bibitem{Sharma2015CR}
S.~K. {Sharma}, T.~E. {Bogale}, S.~{Chatzinotas}, B.~{Ottersten}, L.~B. {Le},
  and X.~{Wang}, ``Cognitive radio techniques under practical imperfections: A
  survey,'' \emph{IEEE Communications Surveys Tutorials}, vol.~17, no.~4, pp.
  1858--1884, Fourthquarter 2015.

\bibitem{Khan2014CA}
Z.~{Khan}, H.~{Ahmadi}, E.~{Hossain}, M.~{Coupechoux}, L.~A. {Dasilva}, and
  J.~J. {Lehtomäki}, ``Carrier aggregation/channel bonding in next generation
  cellular networks: methods and challenges,'' \emph{IEEE Network}, vol.~28,
  no.~6, pp. 34--40, Nov 2014.

\bibitem{Mukherjee2016LAA}
A.~{Mukherjee}, J.~{Cheng}, S.~{Falahati}, H.~{Koorapaty}, D.~H. {Kang},
  R.~{Karaki}, L.~{Falconetti}, and D.~{Larsson}, ``Licensed-assisted access
  lte: coexistence with ieee 802.11 and the evolution toward 5g,'' \emph{IEEE
  Communications Magazine}, vol.~54, no.~6, pp. 50--57, June 2016.

\bibitem{Yang2016advanced}
C.~{Yang}, J.~{Li}, M.~{Guizani}, A.~{Anpalagan}, and M.~{Elkashlan},
  ``Advanced spectrum sharing in 5g cognitive heterogeneous networks,''
  \emph{IEEE Wireless Communications}, vol.~23, no.~2, pp. 94--101, April 2016.

\bibitem{Bajaj2015spectrumtrading}
I.~{Bajaj}, Y.~H. {Lee}, and Y.~{Gong}, ``A spectrum trading scheme for
  licensed user incentives,'' \emph{IEEE Transactions on Communications},
  vol.~63, no.~11, pp. 4026--4036, Nov 2015.

\bibitem{Simeone2008leasing}
O.~{Simeone}, I.~{Stanojev}, S.~{Savazzi}, Y.~{Bar-Ness}, U.~{Spagnolini}, and
  R.~{Pickholtz}, ``Spectrum leasing to cooperating secondary ad hoc
  networks,'' \emph{IEEE Journal on Selected Areas in Communications}, vol.~26,
  no.~1, pp. 203--213, Jan 2008.

\bibitem{Holfeld2016factory}
B.~{Holfeld}, D.~{Wieruch}, T.~{Wirth}, L.~{Thiele}, S.~A. {Ashraf},
  J.~{Huschke}, I.~{Aktas}, and J.~{Ansari}, ``Wireless communication for
  factory automation: an opportunity for lte and 5g systems,'' \emph{IEEE
  Communications Magazine}, vol.~54, no.~6, pp. 36--43, June 2016.

\bibitem{8685776}
G.~{Hampel}, C.~{Li}, and J.~{Li}, ``{5G} ultra-reliable low-latency
  communications in factory automation leveraging licensed and unlicensed
  bands,'' \emph{IEEE Communications Magazine}, vol.~57, no.~5, pp. 117--123,
  May 2019.

\bibitem{8708839}
I.~{Ridwany} and {Iskandar}, ``Forecast of spectrum requirement for mobile
  broadband,'' in \emph{proc. of International Conference on Telecommunication
  Systems, Services, and Applications (TSSA)}, Oct. 2018, pp. 1--5.

\bibitem{KICTIMT2020Korea}
E.-K. Hong, ``Spectrum needs estimate and k-ict plan for imt2020,'' in \emph{5G
  Forum}, May 2017.

\bibitem{Sharma2019mMTC}
S.~K. Sharma and X.~Wang, ``Towards massive machine type communications in
  ultra-dense cellular iot networks: Current issues and machine
  learning-assisted solutions,'' \emph{IEEE Communications Surveys Tutorials},
  pp. 1--1, 2019.

\bibitem{7289481}
F.~{Azmat}, Y.~{Chen}, and N.~{Stocks}, ``Analysis of spectrum occupancy using
  machine learning algorithms,'' \emph{IEEE Trans. Veh. Technol.}, vol.~65,
  no.~9, pp. 6853--6860, Sep. 2016.

\bibitem{Kulin2018endtoend}
M.~Kulin, T.~Kazaz, I.~Moerman, and E.~D. Poorter, ``End-to-end learning from
  spectrum data: A deep learning approach for wireless signal identification in
  spectrum monitoring applications,'' \emph{IEEE Access}, vol.~6, pp.
  18\,484--18\,501, 2018.

\bibitem{SharmaPIMRC2017}
S.~K. {Sharma} and X.~{Wang}, ``Cooperative sensing delay minimization in
  cloud-assisted dsa networks,'' in \emph{2017 IEEE 28th Annual International
  Symposium on Personal, Indoor, and Mobile Radio Communications (PIMRC)}, Oct
  2017, pp. 1--6.

\bibitem{Liglobecomlearning}
L.~Li, X.~He, and H.~Li, ``Learning the spectrum using collaborative filtering
  in directional millimeter wave networks,'' in \emph{GLOBECOM 2017 - 2017 IEEE
  Global Communications Conference}, Dec 2017, pp. 1--7.

\bibitem{patwary2016universal}
M.~N. Patwary, S.~K. Sharma, S.~Chatzinotas, Y.~Chen, M.~A. Maguid, R.~A.
  Alhameed, J.~Noras, and B.~Ottersten, ``Universal intelligent small cell
  (uniscell) for next generation cellular networks,'' \emph{Digital
  Communications and Networks}, vol.~2, no.~4, pp. 167--174, 2016.

\bibitem{Ofcom_location_licensing_2019}
\BIBentryALTinterwordspacing
``Enabling wireless innovation through local licensing: Shared access to
  spectrum supporting mobile technology,'' Ofcom, Report, Jul. 2019. [Online].
  Available:
  \url{https://www.ofcom.org.uk/\_\_data/assets/pdf\_file/0033/157884/enabling-wireless-innovation-through-local-licensing.pdf}
\BIBentrySTDinterwordspacing

\bibitem{Ofcom_location_licensing_2011_ammeded}
\BIBentryALTinterwordspacing
``Wireless telegraphy (mobile spectrum trading) regulations 2011, as amended,''
  Ofcom, Report. [Online]. Available:
  \url{http://www.legislation.gov.uk/uksi/2011/1507/contents}
\BIBentrySTDinterwordspacing

\bibitem{Ofcom_Tarrif_2019_20}
\BIBentryALTinterwordspacing
``Tariff tables 2019/20,'' Ofcom, Report, Mar. 2019. [Online]. Available:
  \url{https://www.ofcom.org.uk/\_\_data/assets/pdf\_file/0032/141899/tariff-tables-2019-20.pdf}
\BIBentrySTDinterwordspacing

\bibitem{NIC_report_2016}
N.~I. Commission, ``{5G} infrastructure requirements in the {UK},'' LS telcom
  UK, Final Report, Version 3.0, 2016.

\end{thebibliography}

\begin{IEEEbiography}[{\includegraphics[width=1in,height=1.25in,clip,keepaspectratio]{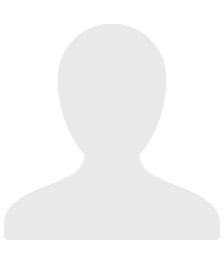}}]{Mohammad N. Patwary}
(SM'11) received the B.Eng. degree (Hons). in electrical and electronic engineering from the Chittagong University of Engineering and Technology, Bangladesh, in 1998, and the Ph.D. degree in telecommunication engineering from The University of New South Wales, Sydney, Australia, in 2005. He was with General Electric Company of Bangladesh from 1998 to 2000 and with Southern Poro Communications, Sydney, from 2001 to 2002, as Research and Development Engineer. He was a Lecturer with The University of New South Wales from 2005 to 2006, and then a Senior Lecturer with Staffordshire University, U.K., from 2006 to 2010. He was then a Full Professor of Wireless Systems and Digital Productivity and the Chair of the Centre of Excellence on Digital Productivity with Connected Services, Staffordshire University, until 2016. He is currently a Full Professor of Telecommunication Networks and Digital Productivity and Head of the Intelligent Systems and Networks (ISN) research group at School of Computing and Digital Technology, Birmingham City University, UK. He is Principal Data Architect for a large scale 5G testbed in the UK to accelerate digital productivity \& develop urban connected community. His current research interests include - sensing and processing for intelligent systems, wireless communication systems design and optimization, signal processing and energy-efficient systems, future generation of cellular network architecture and business modelling for Data-economy.
\end{IEEEbiography}

\begin{IEEEbiography}[{\includegraphics[width=1in,height=1.25in,clip,keepaspectratio]{profile-picture.jpg}}]{Syed Junaid Nawaz} (S'08--M'12--SM'16) received the Ph.D. degree in electronic engineering from Mohammad Ali Jinnah University, Islamabad, in February 2012. Since September 2005, he has worked on several research and teaching positions with COMSATS University Islamabad (CUI), Pakistan; Staffordshire University, UK; Federal Urdu University, Pakistan; The University of York,
UK; and Aristotle University of Thessaloniki, Greece. He is currently working as an Assistant Professor with the Department of Electrical Engineering, COMSATS University Islamabad (CUI), Islamabad, Pakistan.

His current research interests include physical channel modeling, channel estimation and characterization, massive MIMO systems, adaptive signal processing, machine learning, compressed sensing, mmWave channels, airborne internet, underwater communications, Internet of Things, and vehicle-to-vehicle communications.
\end{IEEEbiography}

\begin{IEEEbiography}[{\includegraphics[width=1in,height=1.25in,clip,keepaspectratio]{profile-picture.jpg}}]{Md. Abdur Rahman} (S'02, M'08, SM'13) is an Associate Professor and Chairman of the Department of Cyber Security and Forensic Computing, College of Computer and Cyber Sciences, University of Prince Muqrin (UPM), Madinah Al Munawwarah, Kingdom of Saudi Arabia. Dr. Rahman holds an honorary external research fellowship of King's College London (KCL), UK. His research interests include Blockchain and off-chain solutions, Augmented Reality/Virtual Reality/Mixed Reality based Visualization for Digital Twin systems, Explainable AI and Digital Twin for smart city, Cyber Physical Multimedia Systems, Secure Serious games, Security for Internet of Vehicles, UUV, UAV, EV and AVs, Crowdsourcing and Crowdsensing in smart city, Cloud, fog and multimedia for healthcare Security, IoT and 5G security, Secure Smart City services, Secure ambient intelligent systems, spatio-temporal multimedia big data security, and next generation media security. He is the recipient of BEST Researcher Award from UPM for the year 2018 for winning the prestigious UK-GCC joint collaboration funding through British Council, UK. He has authored more than 105 publications. He has 1 US patent granted and several are pending. Dr. A. Rahman has received more than 12 million SAR as research grant from KACST, KSA and from other international funding bodies. Dr. A. Rahman is the founding director of Advanced Media Laboratory and Smart City Research Laboratory. Dr. Rahman has received three best paper awards from ACM and IEEE Conferences. Dr. Abdur Rahman has successfully supervised 2 PhD students. Dr. Abdur Rahman has published in top tier journals with very high impact factor such as IEEE Communications Magazine, IEEE Internet of Things, Elsevier Future Generation Computer System, IEEE Access, IEEE Transactions on Instrumentation and Measurement, IEEE Sensors, Elsevier Parallel and Distributed Computing, to name a few. Dr. Abdur Rahman is a member of ACM and a senior member of IEEE.
\end{IEEEbiography}

\begin{IEEEbiography}[{\includegraphics[width=1in,height=1.25in,clip,keepaspectratio]{profile-picture.jpg}}] {Shree K. Sharma} (S'12-M'15-SM'18) is currently Research scientist at the SnT, University of Luxembourg. Prior to this, he worked as a Postdoctoral Fellow at the University of Western Ontario, Canada, and also worked as a Research Associate at the SnT being involved in different European, national and ESA projects after receiving his PhD degree in Wireless Communications from the University of Luxembourg in 2014. His current research interests include 5G and beyond wireless, Internet of Things, machine learning, edge computing and optimization of distributed communications, computing and caching resources.

He has published about 100 technical papers in scholarly journals, international conferences and book chapters, and has over 1900 google scholar citations. He is a Senior Member of IEEE and is the recipient of several prestigious awards including ``2018 EURASIP Best Journal Paper Award'', ``Best Paper Award'' in CROWNCOM 2015 conference and ''FNR Award for Outstanding PhD Thesis 2015'' from the FNR, Luxembourg.  He has been serving as a Reviewer for several international journals and conferences; as a TPC member for a number of international conferences including IEEE ICC, IEEE GLOBECOM, IEEE PIMRC, IEEE VTC and IEEE ISWCS; and an Associate Editor for IEEE Access journal. He co-organized a special session in IEEE PIMRC 2017, a workshop in IEEE SECON 2018, worked as a Track co-chair for IEEE VTC-fall 2018 conference, and is a lead editor of the IET book on ``Satellite Communications in the 5G Era''.
\end{IEEEbiography}

\begin{IEEEbiography}[{\includegraphics[width=1in,height=1.25in,clip,keepaspectratio]{profile-picture.jpg}}]{MD MAMUNUR RASHID} is as a Senior Research Fellow at King's Business School, King's College London. He is currently working in a leading Digital Analytics centre called “Consumer and Organizational Digital Analytics Research Centre (CODA)”. Previously, he worked as a Scientific Research Computing Specialist for 3 years in the Department of Engineering Science, at the University of Oxford. Prior to Oxford, he worked 3 years in the Physics Department, at Imperial College London. He was awarded PhD Scholarship to work at the European Organisation for Nuclear Research (CERN), Switzerland after he started his PhD at University of Cranfield. He was awarded a further scholarship by the Atlantic Association for Research in the Mathematical Sciences (AARMS) at the University of New Brunswick, Canada. He also obtained a CERN School of Computing programming diploma from the University of Gottingen, Germany. He worked on Big Data and its analysis of infrastructure development for running various natural disaster impact analysis models to estimate the loss and damage due to the extreme event. His research projects have involved large volume infrastructural data analysis in a distributed HPC environment on the commodity hardware. Coming from a strong infrastructure deployment and data analytical background as well as helping building a number of data-intensive systems, Mamun now works on solving the diverse set of problems for finding impacts of state-of-the-technology in the area of IoT, Big Data, Block Chain, pattern recognition, Smart Infrastructure, Future Cities and Distributed HPC.

In his research career, he has successfully secured a number of scientific research and travel grants from the Natural Environment Research Council (NERC) and Newton Fund (British Council) with Brazil, Thailand, Turkey, Peru, China, Bangladesh, Kazakhstan, Azerbaijan, Dubai, Vietnam and Azerbaijan. Alongside his current position, he is also involved in a number of international multidisciplinary collaborative research activities. He has interests in multi-disciplinary research spectrums focussing on a force for innovation, scientific discovery and potentially those can make a worldwide impact.
\end{IEEEbiography}

\begin{IEEEbiography}[{\includegraphics[width=1in,height=1.25in,clip,keepaspectratio]{profile-picture.jpg}}]{Stuart J. Barnes} is Chaired Professor and Director of the Centre for Consumer and Organisational Digital Analytics (CODA) at King's Business School, King's College London. He joined King's College London in September 2015, having held chair positions at other universities since 2005. A polymath, Stuart is opposed to disciplinary silos and enjoys working across a number of academic disciplines. Stuart reviews for many leading research grant bodies and journals, is associate editor of Information \& Management, and has been involved as programme committee member or track chair in more than 50 academic conferences. Recent research projects have focused on the sharing economy, social media, big data, mobile communications, virtual reality, and virtual worlds. Stuart has published five books (one a bestseller for Butterworth-Heinemann) and more than two hundred articles in leading outlets.
\end{IEEEbiography}

\end{document}